\documentclass{chet}
\usepackage{textcomp}
\usepackage{amssymb}
\usepackage{amsfonts}
\usepackage{graphicx}
\usepackage{subfig}
\preprint{YITP-SB-16-12}
\title{Non-perturbative Four-point Scattering from First-quantized Relativistic JWKB}
\author{M.E. Irizarry-Gelp\'{i}\email{melvineloy@gmail.com} \& W. Siegel\email{siegel@insti.physics.sunysb.edu}}
\affiliation{Chen Ning Yang Institute for Theoretical Physics\\Stony Brook University}
\date{\today}
\abstract{We apply the quantum mechanical (first-quantized) JWKB approximation to a two-body path integral describing the near-forward scattering of two relativistic, heavy, non-identical, scalar particles in $D$ spacetime dimensions. In contrast to the loop expansion, in $D = 4$ this gives a strong-coupling expansion, and in $D = 3$ a non-perturbative weak-coupling expansion.
When the interaction is mediated by massless quanta with spin $N$, we obtain explicit, relativistic results for the scattering amplitude when $N = 0$, $1$ and $2$. In $D = 4$ we find a Regge trajectory function that agrees with the usual quantum mechanical spectrum. We also find an exponentiated infrared divergence that becomes a pure phase factor when the Mandelstam invariants $s$ and $t$ are inside of the physical scattering region. In $D = 3$ we find a singularity whose position along the $s$ axis is dependent on $t$.
When the interaction is mediated by a heavy scalar with mass $M$, in $D = 3$ we find an all-order scattering amplitude where the multi-mass branch points $t = (L + 1)^{2}M^{2}$ appear as Regge poles.}
\begin{document}
\maketitle
\toc
\newpage
\section{Introduction}
Scattering amplitudes are important quantities that bridge theoretical and experimental results. Many tools for the computation of amplitudes have been developed over the past decades. Exact perturbative amplitudes (i.e. tree-level, one-loop, two-loops, etc.) with generic kinematic data can be computed for many theories (see \cite{Elvang:2013cua} for a review). There are also tools for the computation of non-perturbative amplitudes, but these typically involve adding an infinite set of perturbative amplitudes with restricted kinematic data (e.g. Regge limit, fixed-angle limit, etc.).

Some years ago, a program was started by Alday \& Maldacena \cite{Alday:2007hr} to study fixed-angle scattering of gluons in planar $\mathcal{N} = 4$ super Yang-Mills theory at strong coupling. The computation of the scattering amplitude of gluons at strong coupling translates, via the AdS/CFT correspondence, to the computation of a semiclassical amplitude for bosonic strings propagating in $AdS_{5}$. The resulting four-point amplitude at strong coupling agrees with an ansatz of Bern, Dixon \& Smirnov \cite{Bern:2005iz} for the four-point planar MHV scattering amplitude. Other results from this program include dual conformal symmetry, the Yangian and the relation to Wilson loops \cite{Alday:2008yw,Drummond:2010km,Alday:2010kn}.

It is easy to wonder if analogous results can be obtained for less restrictive theories (i.e. other than planar, superconformal gauge theories with string duals). Halpern \& Siegel \cite{HalpernSiegel} found that the (semiclassical) JWKB approximation of certain quantum mechanical systems (e.g. systems of particles) leads to a strong-coupling expansion. In nonrelativistic quantum mechanics, the semiclassical limit of the Feynman path integral gives
\begin{equation}
	\hbar \rightarrow 0: \qquad \int\limits_{\mathbf{x}_{I}}^{\mathbf{x}_{O}} \mathrm{D}\mathbf{q}(t) \, \exp{\left( - \frac{i}{\hbar} S[\mathbf{q}] \right)} \approx \sqrt{\det{(\mathbf{V})}} \exp{\left( - \frac{i}{\hbar} \Sigma \right)};
	\label{FeynmanPath}
\end{equation}
where
\begin{equation}
	\Sigma \equiv S[\mathbf{q}_{*}], \qquad \mathbf{V} \equiv - \frac{i}{\hbar} \frac{\partial^{2} \Sigma}{\partial \mathbf{x}_{I} \partial \mathbf{x}_{O}}.
\end{equation}
Here $\mathbf{q}_{*}(t)$ is a solution of the classical equations of motion with boundary conditions $\mathbf{q}_{*}(t_{I}) = \mathbf{x}_{I}$ and $\mathbf{q}_{*}(t_{O}) = \mathbf{x}_{O}$. The right-hand side of (\ref{FeynmanPath}) is sometimes known as the Van Vleck-Morette approximation to the path integral \cite{VanVleck,CartierMorette}. We use the relativistic sister of this approximation to study four-point near-forward scattering (i.e. small-angle) in a system with two non-identical, heavy, scalar particles. When the particles are coupled via the exchange of massless spin $N$ quanta, in $D$ spacetime dimensions the scattering amplitude takes the form
\begin{equation}
	\mathcal{A} = \delta(P) \left[ \frac{\mathcal{K}_{N}(s)}{\rho_{N}(s)} \right] \int \mathrm{d}B_{12} \, \exp{\left[ - i B_{12} \cdot P_{12} + \beta_{N} \rho_{N}(s) \Gamma(\Delta - 1) \left( \frac{2}{B_{12}^{2}} \right)^{(\Delta - 1)} \right]};
	\label{AmpMassless}
\end{equation}
where $P_{12}^{2} = -t$, $\Delta = (D - 2)/2$, $\beta_{N}$ is a coupling, and the integral over $B_{12}$ is over a volume in $D - 2$ dimensions. This amplitude has the familiar ``eikonal'' form, which was derived long ago by adding perturbative contributions from all ladder diagrams \cite{ChengWuPRL,AbarbItzyk,LevySucher1,ChangMa}.

In \S\ref{sec3} we derive (\ref{AmpMassless}) without any reference to (perturbative) Feynman diagrams in an attempt to make its non-perturbative nature explicit from the beginning. The approach we follow can be viewed as a relativistic generalization of the one used in \cite{ZinnJustinBook} for deriving the nonrelativistic eikonal result for Coulomb scattering. In \S\ref{sec4} and \S\ref{sec5} we evaluate (\ref{AmpMassless}) in $D = 3$ (i.e. $\Delta = 1/2$) and $D = 4$ (i.e. $\Delta = 1$), respectively. In both cases we consider interactions mediated by massless spin $0$, $1$ and $2$ quanta, and we find amplitudes that exhibit bound state singularities. In four spacetime dimensions we find the familiar spectrum with an infinite number of bound states \cite{BIZJ,KabatOrtiz,Dittrich}, but in three spacetime dimensions we find an amplitude with only one bound-state-like singularity, even for the exchange of non-propagating three-dimensional massless spin $2$ quanta.

Then, in \S\ref{sec6} we consider the exchange of massive, spin $0$ quanta in $D = 3$ and $D = 5$ (such that $D - 2 = 1$ and $3$, which are the cases when the long-distance propagator is exact). For the massive exchange in $D = 3$ we find an amplitude that exhibits an infinite number of singularities, but these correspond to the multi-mass branch points (and does not involve the branch cut continuum). Alas, in $D = 5$ it becomes necessary to perform a perturbative expansion in the coupling. We find the expected tree-level amplitude, along with a finite one-loop amplitude and divergent higher-loop amplitudes.

In the next section we begin by introducing the approximation that we use, the forward-JWKB approximation, and contrasting it with another commonly-used approximation, the Regge limit.
\section{Forward-JWKB Approximation\label{sec2}}
In order to be concrete, we consider a four-point scattering event with two non-identical, massive, scalar particles that exchange massless, scalar quanta via a cubic interaction. The $s$-channel process is elastic:
\begin{equation}
	\Phi_{1}(p_{1}) + \Phi_{2}(p_{2}) \longrightarrow \Phi_{1}(p_{3}) + \Phi_{2}(p_{4}).
	\label{ScaProc}
\end{equation}
In this channel, the center-of-momentum energy is given by $\sqrt{s}$ and the magnitude of the momentum transfer is $\sqrt{-t}$. More details about kinematics can be found in Appendix \ref{app1}.
\subsection{Regge Limit}
Consider the one-loop contribution from the box diagram:
\begin{equation}
	\vcenter{\hbox{\includegraphics[scale=0.4]{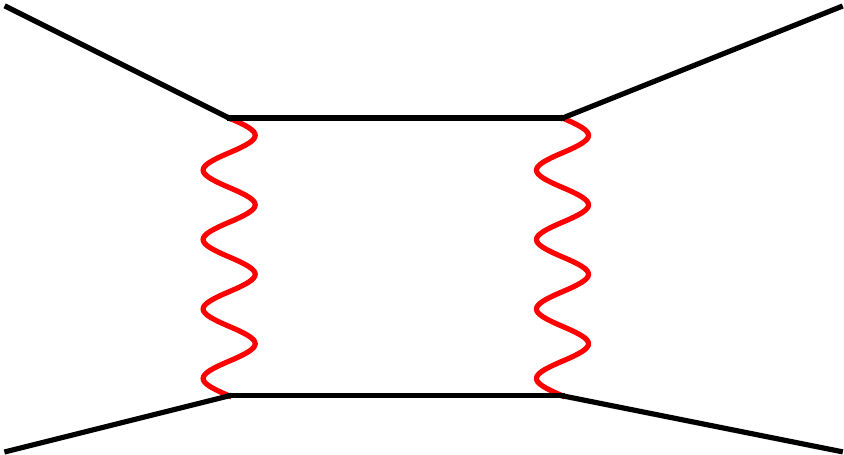}\put(-110,50){$1$} \put(-110,-2){$2$} \put(3,50){$3$} \put(5,-4){$4$}}} \label{BoxDiagram}
\end{equation}
The scattering amplitude from this contribution can be written as an integral over Feynman variables:
\begin{equation}
	\mathcal{A}_{\text{box}}(s, t) \sim g^{4} \Gamma\left( \frac{8 - D}{2} \right) \int\limits_{0}^{1} \int\limits_{0}^{1} \int\limits_{0}^{1} \int\limits_{0}^{1} \mathrm{d}f_{31} \mathrm{d}f_{42} \mathrm{d}f_{12} \mathrm{d}f_{34} \, \frac{\delta(1 - f_{12} - f_{34} - f_{31} - f_{42})}{[B(s, t|f_{ij})]^{(8 - D)/2}};
	\label{BoxIntegral}
\end{equation}
where
\begin{equation}
	B(s, t|f_{ij}) \equiv m_{1}^{2} f_{31}^{2} + m_{2}^{2} f_{42}^{2} + (m_{1}^{2} + m_{2}^{2} - s) f_{31} f_{42} - t f_{12} f_{34}.
\end{equation}
This expression is valid for generic values of $s$ and $t$. However, in the \textbf{Regge limit},
\begin{equation}
	\frac{t}{m_{1} m_{2}} \rightarrow \infty, \qquad \frac{s}{m_{1} m_{2}} \text{ fixed,} \qquad \frac{m_{1}}{m_{2}} \text{ fixed,}
	\label{ReggeLimit}
\end{equation}
the integrals in (\ref{BoxIntegral}) can be evaluated with asymptotic methods \cite{SMatrixBook}. In $D = 4$ one finds
\begin{equation}
	\mathcal{A}_{\text{box}}(s, t) \sim \frac{g^{2}}{t} \left[ g^{2} \rho(s) \right] \log{\left( -\frac{t}{2\mu^{2}} \right)};
\end{equation}
where
\begin{equation}
	\rho(s) \equiv \int\limits_{0}^{1} \frac{\mathrm{d}f}{m_{2}^{2} + (m_{1}^{2} - m_{2}^{2} - s)f + s f^{2}} = \frac{1}{\sqrt{-\Lambda_{12}}} \log{\left[ \frac{s - m_{1}^{2} - m_{2}^{2} + \sqrt{\Lambda_{12}}}{s - m_{1}^{2} - m_{2}^{2} - \sqrt{\Lambda_{12}}} \right]}.
	\label{LeeSawyer}
\end{equation}
with $\Lambda_{12}$ the K\"{a}ll\'{e}n function,
\begin{equation}
	\Lambda_{12} \equiv [s - (m_{1} - m_{2})^{2}][s - (m_{1} + m_{2})^{2}].
\end{equation}
Actually, this result for the one-loop scalar box in the Regge limit agrees with the exact result \cite{PvN,tHVelt}. The logarithm term in (\ref{LeeSawyer}) can be recognized as twice the sum of the rapidities of the incoming states in the center-of-momentum frame (see Appendix \ref{app1}).

Diagrammatically, the Regge limit turns the square polygon in (\ref{BoxDiagram}) into a digon made of matter lines, multiplied by a certain overall factor:
\begin{equation}
	\vcenter{\hbox{\includegraphics[scale=0.4]{box.pdf}}} \quad \longrightarrow \quad (\cdots)\vcenter{\hbox{\includegraphics[scale=0.4]{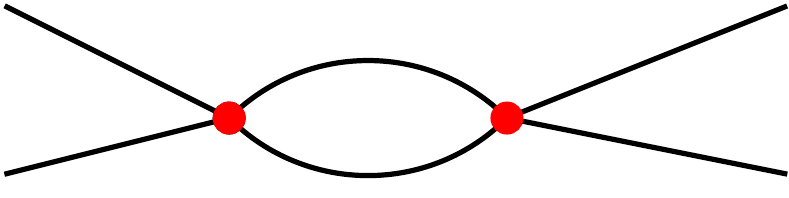}}}
	\label{ReggeBox}
\end{equation}
Whereas the external lines remain unchanged, the internal lines on the left-hand side are matter propagators in $D$ dimensions, while those on the right-hand side are matter propagators in $D-2$ dimensions. Similarly, the double box diagram becomes the concatenation of two matter digons:
\begin{equation}
	\vcenter{\hbox{\includegraphics[scale=0.4]{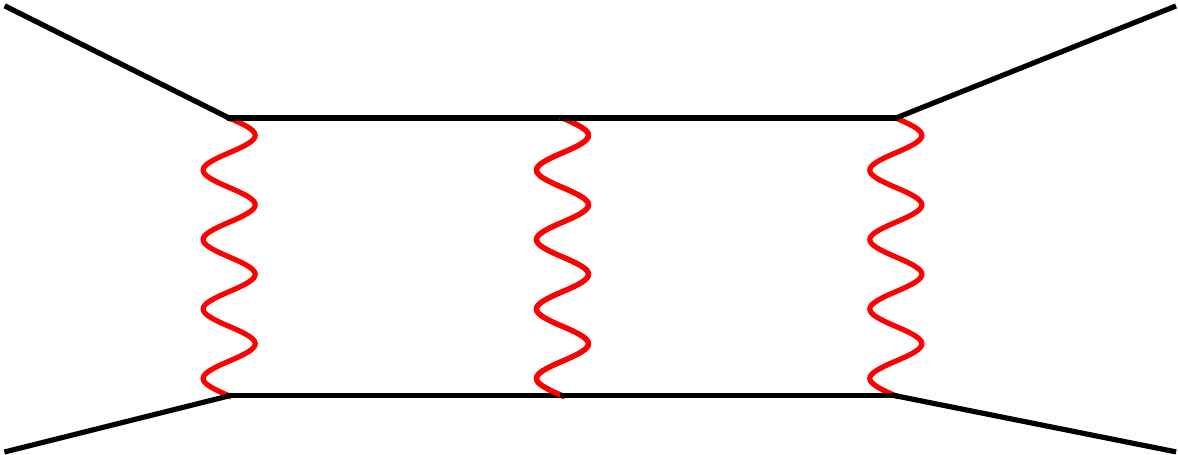}}} \quad \longrightarrow \quad (\cdots)\vcenter{\hbox{\includegraphics[scale=0.4]{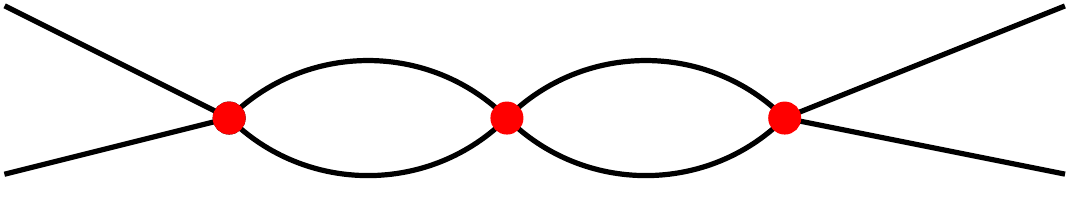}}} \label{ReggeDoubleBox}
\end{equation}
In this way the sum over ladder diagrams becomes much simpler in the Regge limit \cite{LeeSawyer}:
\begin{equation}
	\mathcal{A}_{\text{ladders}}(s, t) \sim g^{2} \left(- \frac{t}{2 \mu^{2}} \right)^{R(s)}, \qquad R(s) = -1 + g^{2} \rho(s) \label{ladders}.
\end{equation}
Note that the Regge limit takes us outside of the physical scattering region of the $s$-channel process. One way to motivate the use of the Regge limit is to consider the (inelastic) $t$-channel process:
\begin{equation}
	\Phi_{1}(p_{1}) + \bar{\Phi}_{1}(\bar{p}_{2}) \longrightarrow \bar{\Phi}_{2}(\bar{p}_{3}) + \Phi_{2}(p_{4}).
\end{equation}
In this channel, the center-of-momentum energy is $\sqrt{t}$ and the magnitude of the momentum transfer is $\sqrt{-s}$ (after crossing from the $s$-channel). Thus, the Regge limit (\ref{ReggeLimit}) in the $s$-channel corresponds to the high-energy and fixed-momentum transfer regime in the $t$-channel. Note that this regime involves \textit{light} masses, or via $\lambda_{j} = \hbar / m_{j}$, large Compton wavelengths. That is, the Regge limit involves distances that are much smaller than the Compton wavelengths (microscopic regime). This is the same regime as taking $\hbar \rightarrow \infty$, so in a way the Regge limit takes us deep into the quantum realm. Indeed, the spectrum of bound states that follows from $R(s) = 0, 1, 2, \ldots$ in (\ref{ladders}) involves the exact one-loop (quantum) result (\ref{LeeSawyer}), and diagrams like (\ref{ReggeDoubleBox}) involve vertices with only internal (quantum) lines attached to them.

Systems where the interaction is mediated by massive quanta can be considered in a similar way. Since the exchange propagators do not appear in the digon ladders, the results should be similar to the massless case. This is a shortcoming of the ladder approach, since massive and massless mediation lead to very different phenomena.
\subsection{Forward-JWKB Regime}
In contrast to the Regge limit (\ref{ReggeLimit}), we \textit{define} the \textbf{forward-JWKB limit} as
\begin{equation}
	{-\frac{t}{m_{1} m_{2}}} \rightarrow 0^{+}, \qquad \frac{s}{m_{1} m_{2}} \text{ fixed,} \qquad \frac{m_{1}}{m_{2}} \text{ fixed.} \label{fJWKBLimit}
\end{equation}
This regime is the same as restricting to small (but physical) scattering angles. Note that the Regge limit corresponds to unphysical scattering angles (i.e. $z_{s} \rightarrow \infty$). From (\ref{fJWKBLimit}) it follows that $-t/s \rightarrow 0^{+}$, meaning that the center-of-momentum energy $\sqrt{s}$ is much larger than the magnitude of the momentum transfer $\sqrt{-t}$. In other words, this is a \textit{high-energy} approximation (in the $s$-channel, the Regge limit is a \textit{low-energy} approximation because $t/s \rightarrow \infty$). Moreover, it also follows that $-t/m_{1}^{2} \rightarrow 0^{+}$ and $-t/m_{2}^{2} \rightarrow 0^{+}$, which mean that the external masses $m_{j}$ are very large compared to $\sqrt{-t}$. Thus, this regime involves \textit{heavy} masses, or via $\lambda_{j} = \hbar / m_{j}$, short Compton wavelengths. Hence, the forward-JWKB limit involves distances that are much larger than the Compton wavelengths (macroscopic regime), which coincides with the limit $\hbar \rightarrow 0$: the (semiclassical) JWKB limit.

Although we are not going to use perturbative second-quantized Feynman diagrams, it is somewhat illuminating to see what happens to the ladder series in the forward-JWKB limit. In contrast to (\ref{ReggeBox}), the one-loop box becomes a digon made with mediator lines,
\begin{equation}
	\vcenter{\hbox{\includegraphics[scale=0.4]{box.pdf}}} \quad \longrightarrow \quad (\cdots)\vcenter{\hbox{\includegraphics[scale=0.4]{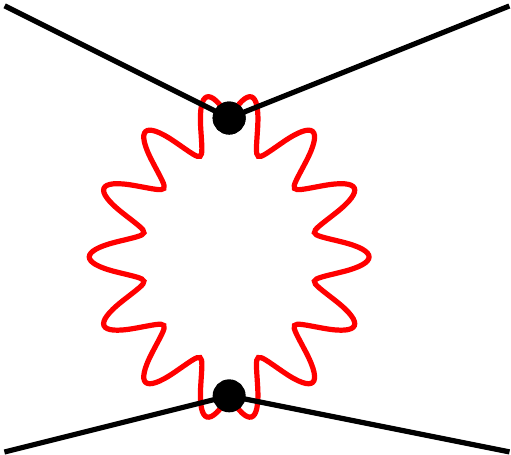}}} \label{fJWKBBox}
\end{equation}
and similarly for the two-loop double box:
\begin{equation}
	\vcenter{\hbox{\includegraphics[scale=0.4]{doublebox.pdf}}} \quad \longrightarrow \quad (\cdots)\vcenter{\hbox{\includegraphics[scale=0.4]{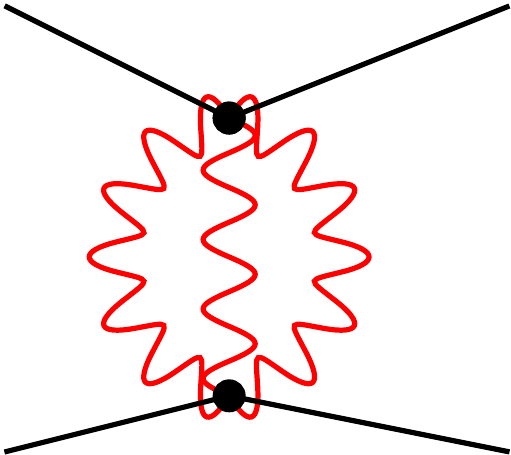}}} \label{fJWKBDoubleBox}
\end{equation}
We see that the structure of the forward-JWKB ladders is very different from the Regge ladders. Indeed, the forward-JWKB ladders do not involve any internal vertices and only involve internal mediator lines (which, like the internal matter lines in the Regge ladders, live in $D - 2$ dimensions). At face value, the forward-JWKB approximation does not make the evaluation of the integrals in (\ref{BoxIntegral}) any easier. This is why we do not explicitly prove (\ref{fJWKBBox}) and (\ref{fJWKBDoubleBox}) for generic theories. However, in \S\ref{sec6} we obtain a scattering amplitude in $D = 3$ using the forward-JWKB approximation that agrees with the expectations from (\ref{fJWKBBox}), (\ref{fJWKBDoubleBox}) and beyond.
\section{Path Integrals\label{sec3}}
We treat the matter quanta as particles (i.e. not fields). In the models that we study, each particle couples to a mediating field $H_{N}$ with spin $N \geq 0$. We will consider massless mediating fields with $N = 0$, $1$ and $2$, and a massive mediating field with $N = 0$. The action functional for the $\Phi_{1}$ and $\Phi_{2}$ particles in (\ref{ScaProc}) has the form
\begin{equation}
	S_{\text{part}}[ q_{1}, q_{2}, H_{N} ] = S_{\text{free}}[ q_{1}, q_{2} ] + S_{\text{int}}[ q_{1}, q_{2}, H_{N} ].
	\label{SP}
\end{equation}
Here, $S_{\text{free}}$ contains the (free) worldline-gauge-fixed kinetic terms,
\begin{equation}
	S_{\text{free}}[ q_{1}, q_{2} ] = \frac{1}{2}\int \mathrm{d}\tau_{1} \left[-\dot{q}_{1}^{2} + M_{1}^{2} \right] + \frac{1}{2} \int \mathrm{d}\tau_{2} \left[- \dot{q}_{2}^{2} + M_{2}^{2} \right];
\end{equation}
and $S_{\text{int}}$ contains the worldline-gauge-fixed terms with the coupling to the field $H_{N}$,
\begin{equation}
\begin{split}
	S_{\text{int}}[ q_{1}, q_{2}, H_{N} ] = {}& \frac{g_{N}}{N!} \int \mathrm{d} \tau_{1} \, \dot{q}_{1}^{a_{1}} \cdots \dot{q}_{1}^{a_{N}} (H_{N}[q_{1}(\tau_{1})])_{a_{1} \cdots a_{N}} \\
	&+ \frac{g_{N}}{N!} \int \mathrm{d} \tau_{2} \, \dot{q}_{2}^{b_{1}} \cdots \dot{q}_{2}^{b_{N}} (H_{N}[q_{2}(\tau_{2})])_{b_{1} \cdots b_{N}};
\end{split}
\label{Sint}
\end{equation}
where $g_{N}$ is a dimensionful coupling, and the $M_{i}$ are ``internal'' worldline masses which are a priori different from the external masses $m_{i}$. The field $H_{N}$ is totally symmetric in the $N$ spacetime indices.

The mediating field $H_{N}$ is made dynamical by adding a kinetic term to the particle action (\ref{SP}),
\begin{equation}
	S[ q_{1}, q_{2}, H_{N} ] = S_{\text{kin}}[H_{N}] + S_{\text{part}}[ q_{1}, q_{2}, H_{N} ].
\end{equation}
We will mostly consider the free, massless case. That is, when $N = 0$ we have a free, massless, scalar field $H_{0}$; when $N = 1$ we have an abelian, vector field $(H_{1})_{a}$ (photon); and when $N = 2$ we have a linearized, symmetric, tensor field $(H_{2})_{ab}$ (massless Fierz-Pauli graviton). After an appropriate gauge-fixing (we use the analog of the Fermi-Feynman gauge), the kinetic term in $S_{\text{kin}}$ takes the form
\begin{equation}
	S_{\text{kin}}[H_{N}] = \frac{1}{2} \int \int \mathrm{d}x \mathrm{d}y \left( [H_{N}(x)]_{a_{1} \cdots a_{N}} [K_{N}(x|y)]^{a_{1} \cdots a_{N} b_{1} \cdots b_{N}} [H_{N}(y)]_{b_{1} \cdots b_{N}} \right);
\end{equation}
where the free, massless, spin $N$ gauge-fixed kinetic operator $K_{N}$ is given by
\begin{equation}
	[K_{N}(x|y)]^{a_{1} \cdots a_{N} b_{1} \cdots b_{N}} = (\kappa_{N})^{a_{1} \cdots a_{N} b_{1} \cdots b_{N}} K_{0}(x|y);
\end{equation}
with $\kappa_{N}$ a constant tensor that is separately totally symmetric in the $a_{j}$ and $b_{k}$ indices, and $K_{0}$ is the free, massless, scalar kinetic operator,
\begin{equation}
	K_{0}(x|y) = \delta(x - y) \left(- \frac{1}{2} \partial^{2} \right).
\end{equation}
\subsection{Forward-JWKB Path Integrals}
The system $H_{N} + \Phi_{1} + \Phi_{2}$ is described by the un-integrated quantum path integral (i.e. a path integral that is dependent on the modulus of each worldline):
\begin{equation}
	\mathcal{F}_{T}(3, 4|1, 2) = \int \widehat{\mathrm{D}}H_{N}(x) \int\limits_{x_{1}}^{x_{3}} \mathrm{D}q_{1}(\tau_{1}) \int\limits_{x_{2}}^{x_{4}} \mathrm{D}q_{2}(\tau_{2}) \exp{\left( - i S\left[ q_{1}, q_{2}, H_{N} \right] \right)}.
\end{equation}
The functional measure over $H_{N}$ is normalized such that
\begin{equation}
	\int \widehat{\mathrm{D}}H_{N}(x) \exp{\left(- i S_{\text{kin}}[H_{N}] \right)} = 1.
\end{equation}
That is, if $g_{N} = 0$, then $\mathcal{F}_{T}$ reduces to the product of two un-integrated free, massive, scalar Green functions.

After writing $S_{\text{int}}$ in (\ref{Sint}) as a spacetime volume integral,
\begin{equation}
	S_{\text{int}}[ q_{1}, q_{2}, H_{N} ] = \int \mathrm{d}x \left[ J_{1}(x) \cdot H_{N}(x) + J_{2}(x) \cdot H_{N}(x) \right];
\end{equation}
with the aid of sources $J_{1}$ and $J_{2}$ given by
\begin{align}
	[J_{1}(x)]^{a_{1} \cdots a_{N}} &\equiv \frac{g_{N}}{N!} \int \mathrm{d} \tau_{1} \, \dot{q}_{1}^{a_{1}} \cdots \dot{q}_{1}^{a_{N}} \delta[x - q_{1}(\tau_{1})], \\
	[J_{2}(x)]^{b_{1} \cdots b_{N}} &\equiv \frac{g_{N}}{N!} \int \mathrm{d} \tau_{2} \, \dot{q}_{2}^{b_{1}} \cdots \dot{q}_{2}^{b_{N}} \delta[x - q_{2}(\tau_{2})];
\end{align}
we perform the functional integral over $H_{N}$ and find
\begin{equation}
	\mathcal{F}_{T}(3, 4|1, 2) = \int\limits_{x_{1}}^{x_{3}} \mathrm{D}q_{1}(\tau_{1}) \int\limits_{x_{2}}^{x_{4}} \mathrm{D}q_{2}(\tau_{2}) \exp{\left( - i S_{\text{eff}}\left[ q_{1}, q_{2}\right] \right)};
	\label{FSeff}
\end{equation}
with the functional $S_{\text{eff}}$ given by
\begin{equation}
	S_{\text{eff}}\left[ q_{1}, q_{2}\right] \equiv S_{\text{free}}\left[ q_{1}, q_{2}\right] - \frac{1}{2} \int \int \mathrm{d}x \mathrm{d}y \left[ J_{1}(x) + J_{2}(x) \right] \cdot G_{N}(x|y) \cdot \left[ J_{1}(y) + J_{2}(y) \right].
	\label{Seff}
\end{equation}
Here $G_{N} = (K_{N})^{-1}$ is the free, massless, spin $N$ gauge-fixed Green function,
\begin{equation}
	[G_{N}(x|y)]_{a_{1} \cdots a_{N} b_{1} \cdots b_{N}} = -i (\nu_{N})_{a_{1} \cdots a_{N} b_{1} \cdots b_{N}} \Gamma\left( \Delta \right) \left[ \frac{2}{(x - y)^{2}} \right]^{\Delta}, \qquad \Delta \equiv \frac{D - 2}{2};
	\label{GN}
\end{equation}
with $\nu_{N}$ a constant tensor that satisfies
\begin{equation}
	(\nu_{N})_{a_{1} \cdots a_{N} c_{1} \cdots c_{N}} (\kappa_{N})^{c_{1} \cdots c_{N} b_{1} \cdots b_{N}} = \frac{1}{N!} \left( \delta_{a_{1}}{}^{b_{1}} \cdots \delta_{a_{N}}{}^{b_{N}} + \text{ permutations} \right).
\end{equation}

The outcome of integrating over $H_{N}$ is the appearance of ``potential'' terms that describe the effective interactions between the $\Phi_{1}$ and $\Phi_{2}$ particles. Indeed, we write $S_{\text{eff}}$ in (\ref{Seff}) as
\begin{equation}
	S_{\text{eff}}[q_{1}, q_{2}] = S_{\text{free}}[q_{1}, q_{2}] - S_{\text{one}}[q_{1}, q_{2}] - S_{\text{two}}[q_{1}, q_{2}];
\end{equation}
with $S_{\text{one}}$ containing self-interaction terms,
\begin{equation}
\begin{split}
	S_{\text{one}} = {}& \frac{g_{N}^{2}}{2 (N!)^{2}} \int \int \mathrm{d}\tau_{1} \mathrm{d}\sigma_{1} \left[ \dot{q}_{1}(\tau_{1}) \cdots \dot{q}_{1} \cdot G_{N}[q_{1}(\tau_{1}) | q_{1}(\sigma_{1})] \cdot \dot{q}_{1}(\sigma_{1}) \cdots \dot{q}_{1} \right] \\
	&+ \frac{g_{N}^{2}}{2 (N!)^{2}} \int \int \mathrm{d}\tau_{2} \mathrm{d}\sigma_{2} \left[ \dot{q}_{2}(\tau_{2}) \cdots \dot{q}_{2} \cdot G_{N}[q_{2}(\tau_{2}) | q_{2}(\sigma_{2})] \cdot \dot{q}_{2}(\sigma_{2}) \cdots \dot{q}_{2} \right];
\end{split}
\label{S10}
\end{equation}
and $S_{\text{two}}$ containing a two-body interaction term,
\begin{equation}
	S_{\text{two}} = \frac{g_{N}^{2}}{(N!)^{2}} \int \int \mathrm{d}\tau_{1} \mathrm{d}\tau_{2} \left[ \dot{q}_{1}(\tau_{1}) \cdots \dot{q}_{1} \cdot G_{N}[q_{1}(\tau_{1}) | q_{2}(\tau_{2})] \cdot \dot{q}_{2}(\tau_{2}) \cdots \dot{q}_{2} \right].
	\label{S20}
\end{equation}
One can think of the first term in $S_{\text{one}}$ as summing over contributions that involve linking a point $q_{1}(\tau_{1})$ to a point $q_{1}(\sigma_{1})$ with a propagator $G_{N}$. Both of these points are on the $\Phi_{1}$ worldline. Similarly, the second term in $S_{\text{one}}$ involves linking two points on the $\Phi_{2}$ worldline. On the other hand, $S_{\text{two}}$ can be understood as summing over contributions that involve linking a point $q_{1}(\tau_{1})$ on the $\Phi_{1}$ worldline to a point $q_{2}(\tau_{2})$ on the $\Phi_{2}$ worldline.

At this stage, the discussion is general and exact. In what follows \textbf{we ignore the self-interaction terms}. When this is done, $\mathcal{F}_{T}$ in (\ref{FSeff}) takes the form
\begin{equation}
	\mathcal{F}_{T}(3, 4|1, 2) = \int\limits_{x_{1}}^{x_{3}} \mathrm{D}q_{1}(\tau_{1}) \int\limits_{x_{2}}^{x_{4}} \mathrm{D}q_{2}(\tau_{2}) \exp{\left( - i S_{\text{free}} \right)} \exp{\left( i S_{\text{two}} \right)}.
	\label{320}
\end{equation}
We \textit{could} expand the second exponential in (\ref{320}) as a perturbative expansion in $g_{N}$,
\begin{equation}
	\exp{\left( i S_{\text{two}} \right)} = 1 + i S_{\text{two}} + \frac{1}{2!} (i S_{\text{two}})^{2} + \frac{1}{3!} (i S_{\text{two}})^{3} + \cdots;
	\label{321}
\end{equation}
and evaluate each contribution. The first term in (\ref{321}) is of order $g_{N}^{0}$ and not very interesting. The second term is proportional to $S_{\text{two}}$ and thus of order $g_{N}^{2}$. It involves a sum over all possible ways to connect a point on the $\Phi_{1}$ worldline to a point on the $\Phi_{2}$ worldline with a $G_{N}$ propagator. This is a tree-level contribution. The third term involves
\begin{equation}
	(i S_{\text{two}})^{2} \sim g_{N}^{4} \int \int \mathrm{d}\tau_{1} \mathrm{d}\tau_{2} \int \int \mathrm{d}\sigma_{1} \mathrm{d}\sigma_{2} (\cdots) G_{N}[q_{1}(\tau_{1}) | q_{2}(\tau_{2})] G_{N}[q_{1}(\sigma_{1}) | q_{2}(\sigma_{2})].
\end{equation}
This can be identified with a one-loop contribution, but since the double integration scans all possible orderings of the worldline coordinates, it accounts for both box-like and crossed box-like contributions. Similarly, the fourth term in (\ref{321}) corresponds to the two-loops contribution, and contains the double box, the crossed double box and other non-planar contributions. Thus, we have learned that $\mathcal{F}_{T}$ in (\ref{320}) contains all perturbative contributions arising from generalized ladder diagrams. Also, it follows that the contributions to the scattering amplitude from $\mathcal{F}_{T}$ are un-truncated and we need to perform some truncation.

We now incorporate the forward-JWKB approximation into our analysis. Since the forward-JWKB approximation is a combination of the forward approximation and the semiclassical approximation, in the forward-JWKB approximation the un-integrated quantum path integral $\mathcal{F}_{T}$ takes the form
\begin{equation}
	\mathcal{F}_{T}(3, 4|1, 2) \longrightarrow \mathcal{G}_{T}(3, 4|1, 2) = \sqrt{-\det{(V)}} \exp{\left( -i \Sigma \right)}.
\end{equation}
where the function $\Sigma$ is the value of $S_{\text{eff}}$ evaluated at the forward paths $f_{1}$ and $f_{2}$,
\begin{equation}
	\Sigma \equiv S_{\text{eff}} [f_{1}, f_{2}];
\end{equation}
and the matrix $V$ is given by
\begin{equation}
	V \equiv \begin{pmatrix}
	V_{1 3} & V_{2 3} \\
	V_{1 4} & V_{2 4}
	\end{pmatrix}, \qquad V_{jk} \equiv - i \frac{\partial \Sigma}{\partial x_{j} \partial x_{k}}.
\end{equation}
The forward paths describe particles that are moving along straight paths in spacetime with fixed spacetime speed:
\begin{equation}
\begin{split}
	f_{1}(\tau_{1}) &= \frac{x_{1} + x_{3}}{2} + \left( \frac{\tau_{1}}{T_{1}} \right) \left(x_{3} - x_{1} \right), \qquad { - \frac{T_{1}}{2} } < \tau_{1} < \frac{T_{1}}{2}; \\
	f_{2}(\tau_{2}) &= \frac{x_{2} + x_{4}}{2} + \left( \frac{\tau_{2}}{T_{2}} \right) \left(x_{4} - x_{2} \right), \qquad { - \frac{T_{2}}{2} } < \tau_{2} < \frac{T_{2}}{2}.
\end{split}
\label{fpaths}
\end{equation}
Here $T_{1}$ and $T_{2}$ are the moduli of the $\Phi_{1}$ and $\Phi_{2}$ worldlines, respectively. The form of $\mathcal{G}_{T}$ can be recognized as the (relativistic) two-body generalization of the Van Vleck-Morette kernel \cite{VanVleck,CartierMorette} specialized to the forward paths (i.e. we do not solve for the true classical paths). For this reason we refer to $\mathcal{G}_{T}$ as the un-integrated forward-JWKB kernel. In the rest of this section we evaluate $\mathcal{G}_{T}$ and relate it to the scattering amplitude.
\subsubsection{Forward Van Vleck Function}
At the forward paths (\ref{fpaths}), the free part of $S_{\text{eff}}$ gives
\begin{equation}
	\Sigma_{\text{free}} \equiv S_{\text{free}}[f_{1}, f_{2}] = - \frac{1}{2 T_{1}} x_{3 1}^{2} + \frac{M_{1}^{2} T_{1}}{2} - \frac{1}{2 T_{2}} x_{4 2}^{2} + \frac{M_{2}^{2} T_{2}}{2}, \qquad x_{jk} \equiv x_{j} - x_{k}.
	\label{SigFree}
\end{equation}
Recall that we are dropping the self-interactions and keeping the two-body interaction. Evaluating $S_{\text{two}}$ at the forward paths gives $\Sigma_{\text{two}} \equiv S_{\text{two}}[f_{1}, f_{2}]$, i.e. 
\begin{equation}
	\Sigma_{\text{two}} = \frac{g_{N}^{2}}{(N!)^{2}} \int \int \mathrm{d}\tau_{1} \mathrm{d}\tau_{2} \left[ \dot{f}_{1}(\tau_{1}) \cdots \dot{f}_{1} \cdot G_{N}[f_{1}(\tau_{1}) | f_{2}(\tau_{2})] \cdot \dot{f}_{2}(\tau_{2}) \cdots \dot{f}_{2} \right].
\end{equation}
Note that the forward paths have constant slope:
\begin{equation}
	\dot{f}_{1} = \frac{x_{31}}{T_{1}} \equiv k_{31}, \qquad \dot{f}_{2} = \frac{x_{42}}{T_{2}} \equiv k_{42}.
	\label{slope}
\end{equation}
Using (\ref{GN}) and (\ref{slope}), we write
\begin{equation}
	\dot{f}_{1}(\tau_{1}) \cdots \dot{f}_{1} \cdot G_{N}[f_{1}(\tau_{1}) | f_{2}(\tau_{2})] \cdot \dot{f}_{2}(\tau_{2}) \cdots \dot{f}_{2} = \mathcal{K}_{N} G_{0}[f_{1}(\tau_{1}) | f_{2}(\tau_{2})];
\end{equation}
where $G_{0}$ is the massless, scalar Green function and the constant, scalar factor $\mathcal{K}_{N}$ consists of contractions of $N$ copies of $k_{3 1}$ and $k_{4 2}$ with the constant tensor $\nu_{N}$:
\begin{equation}
	\mathcal{K}_{N} = \left( \frac{1}{T_{1} T_{2}} \right)^{N} \left(x_{3 1} \cdots x_{3 1} \right) \cdot \nu_{N} \cdot \left(x_{4 2} \cdots x_{4 2}\right) = \left(k_{3 1} \cdots k_{3 1} \right) \cdot \nu_{N} \cdot \left(k_{4 2} \cdots k_{4 2}\right).
\end{equation}
With the forward paths (\ref{fpaths}), we have
\begin{equation}
	f_{1}(\tau_{1}) - f_{2}(\tau_{2}) = X_{1 2} + \left( \frac{\tau_{1}}{T_{1}} \right) x_{3 1} - \left( \frac{\tau_{2}}{T_{2}} \right) x_{4 2};
\end{equation}
where we have introduced
\begin{equation}
	X_{1 2} \equiv \frac{x_{1} - x_{2} + x_{3} - x_{4}}{2}.
\end{equation}
Note that $X_{1 2}$ is the vector average of the separation of the incoming particles (given by the vector $x_{1} - x_{2}$) and the separation of the outgoing particles (given by the vector $x_{3} - x_{4}$). That is, $X_{12}$ is incoming/outgoing symmetric. We can now write $\Sigma_{\text{two}}$ as
\begin{equation}
	\Sigma_{\text{two}} = -i\frac{g_{N}^{2}}{(N!)^{2}} T_{1} T_{2} \mathcal{K}_{N} \Upsilon_{\text{two}};
\end{equation}
with
\begin{equation}
	\Upsilon_{\text{two}} \equiv \Gamma(\Delta) \int\limits_{-1/2}^{1/2}\mathrm{d}u_{1} \int\limits_{-1/2}^{1/2}\mathrm{d}u_{2} \left[ \frac{2}{(X_{1 2} + u_{1} x_{3 1} - u_{2} x_{4 2})^{2}} \right]^{\Delta}, \qquad \Delta \equiv \frac{D - 2}{2}.
\end{equation}
(We have changed variables from $(\tau_{1}, \tau_{2})$ to $(u_{1}, u_{2})$, which are dimensionless). It is convenient to introduce a Schwinger parameter $T$ and write
\begin{equation}
	\Upsilon_{\text{two}} = \int\limits_{-1/2}^{1/2}\mathrm{d}u_{1} \int\limits_{-1/2}^{1/2}\mathrm{d}u_{2} \int\limits_{0}^{\infty}\mathrm{d}T \left( \frac{1}{T} \right)^{(\Delta + 1)} \exp{\left[- \frac{1}{2 T} (X_{1 2} + u_{1} x_{3 1} - u_{2} x_{4 2})^{2} \right]}.
	\label{UpSchw}
\end{equation}
The path difference $f_{1}(u_{1}) - f_{2}(u_{2}) = X_{1 2} + u_{1} x_{3 1} - u_{2} x_{4 2}$ describes the separation between the particles during the scattering process. The conjugate momentum to this separation is the momentum transfer. In the forward-JWKB approximation, the momentum transfer is very small compared to the masses or the center-of-momentum energy. By Fourier-Heisenberg conjugacy, this means that the separation between the particles is always very large compared to the displacement of each particle. Thus, in the forward-JWKB approximation the $(u_{1}, u_{2})$ integral in (\ref{UpSchw}) is dominated by the contribution from the critical point of the expression in the exponent:
\begin{align}
	\bar{u}_{1} &= - \left[ \frac{x_{42}^{2} (X_{12} \cdot x_{31}) - (X_{12} \cdot x_{42})(x_{31} \cdot x_{42})}{x_{31}^{2} x_{42}^{2} - (x_{31} \cdot x_{42})^{2}} \right], \\
	\bar{u}_{2} &= + \left[ \frac{x_{31}^{2} (X_{12} \cdot x_{42}) - (X_{12} \cdot x_{31})(x_{31} \cdot x_{42})}{x_{31}^{2} x_{42}^{2} - (x_{31} \cdot x_{42})^{2}} \right].
\end{align}
At this critical point, we find
\begin{equation}
	B_{12} \equiv f_{1}(\bar{u}_{1}) - f_{2}(\bar{u}_{2}) = X_{1 2} + \bar{u}_{1} x_{3 1} - \bar{u}_{2} x_{4 2};
\end{equation}
which satisfies $B_{12} \cdot x_{31} = 0$ and $B_{12} \cdot x_{42} = 0$. That is, $B_{12}$ is the projection of $X_{12}$ to the subspace that is orthogonal to $x_{31}$ and $x_{42}$. In the forward-JWKB approximation we find
\begin{align}
	\Upsilon_{\text{two}} &\approx \frac{2 \pi}{\sqrt{x_{31}^{2} x_{42}^{2} - (x_{31} \cdot x_{42})^{2}}} \int\limits_{0}^{\infty}\mathrm{d}T \left( \frac{1}{T} \right)^{\Delta} \exp{\left[- \frac{1}{2 T} B_{1 2}^{2} \right]} \nonumber \\
	&= \frac{2 \pi}{\sqrt{x_{31}^{2} x_{42}^{2} - (x_{31} \cdot x_{42})^{2}}} \Gamma(\Delta - 1) \left( \frac{2}{B_{1 2}^{2}} \right)^{(\Delta - 1)};
\end{align}
and hence $\Sigma_{\text{two}}$ gives
\begin{equation}
	\Sigma_{\text{two}} \approx -i \left( \frac{2 \pi g_{N}^{2}}{(N!)^{2}} \right) \left[ \frac{T_{1} T_{2} \mathcal{K}_{N}}{\sqrt{x_{31}^{2} x_{42}^{2} - (x_{31} \cdot x_{42})^{2}}} \right] \Gamma(\Delta - 1) \left( \frac{2}{B_{1 2}^{2}} \right)^{(\Delta - 1)}.
\end{equation}
Note that this expression is divergent when $\Delta = 1$ (i.e. when $D = 4$). In order to keep things compact, we introduce the coupling
\begin{equation}
	\beta_{N} \equiv \frac{2 \pi g_{N}^{2}}{(N!)^{2}};
	\label{betaN}
\end{equation}
and the function
\begin{equation}
	\rho_{N} \equiv \frac{T_{1} T_{2} \mathcal{K}_{N}}{\sqrt{x_{31}^{2} x_{42}^{2} - (x_{31} \cdot x_{42})^{2}}} = \frac{\mathcal{K}_{N}}{\sqrt{k_{31}^{2} k_{42}^{2} - (k_{31} \cdot k_{42})^{2}}}.
\end{equation}
When $\rho_{N}$ is written in terms of $k_{31}$ and $k_{42}$, there are no explicit factors of $T_{1}$ and $T_{2}$.
\subsubsection{Forward Van Vleck Matrix}
Since the Van Vleck function $\Sigma$ has the form $\Sigma_{\text{free}} - \Sigma_{\text{two}}$, we write the Van Vleck matrix $V$ also in the form $V_{\text{free}} - V_{\text{two}}$ with
\begin{equation}
	V_{\text{free}} = \begin{pmatrix}
	u_{1 3} & u_{2 3} \\
	u_{1 4} & u_{2 4}
	\end{pmatrix}, \quad u_{jk} \equiv - i \frac{\partial \Sigma_{\text{free}}}{\partial x_{j} \partial x_{k}}; \quad V_{\text{two}} = \begin{pmatrix}
	v_{1 3} & v_{2 3} \\
	v_{1 4} & v_{2 4}
	\end{pmatrix}, \quad v_{jk} \equiv i \frac{\partial \Sigma_{\text{two}}}{\partial x_{j} \partial x_{k}}.
\end{equation}
The determinant of $V$ can be written as
\begin{equation}
	\det{(V)} = \det(V_{\text{free}} - V_{\text{two}}) = \det{(I - W)} \det{(V_{\text{free}})}, \qquad W \equiv V_{\text{two}} \cdot (V_{\text{free}})^{-1}.
\end{equation}
Hence, the square root of the determinant is
\begin{equation}
	\sqrt{-\det{(V)}} = \sqrt{-\det{(V_{\text{free}})}} \exp{\left[ - \sum_{n = 1}^{\infty} \frac{1}{2 n} \operatorname{tr}{(W^{n})} \right]}.
	\label{sqrtV}
\end{equation}
Using $\Sigma_{\text{free}}$ from (\ref{SigFree}), it is easy to show that
\begin{equation}
\begin{split}
	(u_{13})_{ab} = \left( - \frac{i}{T_{1}} \right) \eta_{ab}, &\qquad (u_{23})_{ab} = 0; \\
	(u_{14})_{ab} = 0, &\qquad (u_{24})_{ab} = \left( - \frac{i}{T_{2}} \right) \eta_{ab};
\end{split}
\end{equation}
and thus
\begin{equation}
	\sqrt{-\det{(V_{\text{free}})}} = \left( - \frac{i}{T_{1}} \right)^{D/2} \left( - \frac{i}{T_{2}} \right)^{D/2}.
\end{equation}
We can think of the terms with traces of $W$ in (\ref{sqrtV}) as corrections to $\Sigma$, since they appear inside an exponential too. If we compute them, we find that they involve powers of $B_{12}^{2}$ that are more negative than the power in $\Sigma_{\text{two}}$. In the forward-JWKB approximation $B_{1 2}^{2}$ is large, so we keep the dominant contribution from $\Sigma_{\text{two}}$ and drop all terms with traces of $W$. Hence,
\begin{equation}
	\sqrt{-\det{(V)}} \approx \sqrt{-\det{(V_{\text{free}})}}.
\end{equation}
This step might seem drastic, but as we shall see, the end result will justify our means.
\subsection{Integrated Forward-JWKB Scattering Kernel}
From $\mathcal{F}_{T}$ we obtain the un-integrated quantum scattering kernel $\mathcal{S}_{T}$ via
\begin{equation}
	\mathcal{S}_{T}(3, 4|1, 2) = \int \int \int \int \mathrm{d}x_{1} \mathrm{d}x_{2} \mathrm{d}x_{3} \mathrm{d}x_{4} \, \overline{\mathcal{W}}_{O}(3, 4) \mathcal{W}_{I}(1, 2) \mathcal{F}_{T}(3, 4|1, 2).
	\label{STTF}
\end{equation}
The factors $\mathcal{W}_{I}$ and $\overline{\mathcal{W}}_{O}$ account for the asymptotic free massive external states:
\begin{align}
	\mathcal{W}_{I}(1, 2) &= \exp{\left[ \frac{i T_{1}}{4} \left( p_{1}^{2} + m_{1}^{2} \right) + \frac{i T_{2}}{4} \left( p_{2}^{2} + m_{2}^{2} \right) + i x_{1} \cdot p_{1} + i x_{2} \cdot p_{2} \right]}; \\
	\overline{\mathcal{W}}_{O}(3, 4) &= \exp{\left[ \frac{i T_{1}}{4} \left( p_{3}^{2} + m_{3}^{2} \right) + \frac{i T_{2}}{4} \left( p_{4}^{2} + m_{4}^{2} \right) - i x_{3} \cdot p_{3} - i x_{4} \cdot p_{4} \right]}.
\end{align}
Note that a priori we have $m_{1} \neq m_{3} \neq M_{1}$ and $m_{2} \neq m_{4} \neq M_{2}$. That is, the external states are off-shell and the external masses $m_{i}$ are not related to the worldline masses $M_{i}$.

In the forward-JWKB approximation, we use $\mathcal{G}_{T}$ instead of $\mathcal{F}_{T}$ in (\ref{STTF}). In order to perform the integration in (\ref{STTF}), we first make a change of position variables and also introduce the corresponding conjugate momenta,
\begin{align}
	X \equiv \frac{x_{1} + x_{2} + x_{3} + x_{4}}{4}, &\qquad P \equiv p_{3} + p_{4} - p_{1} - p_{2}; \\
	X_{12} \equiv \frac{x_{1} - x_{2} + x_{3} - x_{4}}{2}, &\qquad P_{12} \equiv \frac{p_{3} - p_{1} + p_{2} - p_{4}}{2}; \\
	x_{31} \equiv x_{3} - x_{1}, &\qquad p_{31} \equiv \frac{p_{1} + p_{3}}{2};
	\label{p31} \\
	x_{42} \equiv x_{4} - x_{2}, &\qquad p_{42} \equiv \frac{p_{2} + p_{4}}{2};
	\label{p42}
\end{align}
such that
\begin{equation}
	x_{1} \cdot p_{1} + x_{2} \cdot p_{2} - x_{3} \cdot p_{3} - x_{4} \cdot p_{4} = - X \cdot P - X_{12} \cdot P_{12} - x_{31} \cdot p_{31} - x_{42} \cdot p_{42}.
\end{equation}
The Jacobian from this change of variables is a constant, which we ignore
\begin{equation}
	\mathrm{d}x_{1} \mathrm{d}x_{2} \mathrm{d}x_{3} \mathrm{d}x_{4} \sim \mathrm{d}X \mathrm{d}X_{12} \mathrm{d}x_{31} \mathrm{d}x_{42}.
\end{equation}
In terms of these variables we have
\begin{equation}
\begin{split}
	\overline{\mathcal{W}}_{O}(3, 4) \mathcal{W}_{I}(1, 2) = {}& \exp{\left[ -i X \cdot P -i X_{12} \cdot P_{12} -i x_{31} \cdot p_{31} -i x_{42} \cdot p_{42} \right]} \\
	&\times \exp{\left[ \frac{i T_{1}}{2} p_{31}^{2} + \frac{i T_{1}}{32} \left( 2 P_{1 2} + P \right)^{2} + \frac{i T_{1}}{4} (m_{1}^{2} + m_{3}^{2}) \right]} \\
	&\times \exp{\left[ \frac{i T_{2}}{2} p_{42}^{2} + \frac{i T_{2}}{32} \left( 2 P_{1 2} - P \right)^{2} + \frac{i T_{2}}{4} (m_{2}^{2} + m_{4}^{2}) \right]}.
\end{split}
\end{equation}
Since $\Sigma$ and $V$ have no dependence on $X$, the un-integrated forward-JWKB kernel $\mathcal{G}_{T}$ does not depend on $X$. Thus, the integral over $X$ yields a Dirac delta:
\begin{equation}
	\int \mathrm{d}X \exp{(-i X \cdot P)} = \delta(P).
\end{equation}
This Dirac delta imposes the constraint $P = 0$, which leads to
\begin{equation}
	p_{1} + p_{2} = p_{3} + p_{4}.
\end{equation}
That is, the total external momentum is conserved, as expected from translation invariance. After enforcing $P = 0$, we find
\begin{equation}
	P_{1 2} = p_{3} - p_{1} = p_{2} - p_{4} \quad \Longrightarrow \quad P_{1 2}^{2} = -t.
\end{equation}
Next, we tackle the integration over the $x_{ij}$. The exact integration is nontrivial because of the way that $\rho_{N}$ in $\Sigma_{\text{two}}$ depends on these variables. We make another change of variables:
\begin{equation}
	x_{31} = T_{1} k_{31}, \qquad x_{42} = T_{2} k_{42} \quad \Longrightarrow \quad \mathrm{d}x_{31} \mathrm{d}x_{42} = (T_{1} T_{2})^{D} \mathrm{d}k_{31} \mathrm{d}k_{42}.
\end{equation}
In terms of the $k_{ij}$ we have
\begin{equation}
	\Sigma_{\text{free}} = -\frac{T_{1}}{2} \left(k_{31}^{2} - M_{1}^{2} \right) - \frac{T_{2}}{2} \left( k_{42}^{2} - M_{2}^{2} \right);
\end{equation}
and thus, after integrating over $X$ and enforcing $P = 0$, we find
\begin{equation}
\begin{split}
	\overline{\mathcal{W}}_{O} \mathcal{W}_{I} \exp{(-i \Sigma_{\text{free}})} = {}& \exp{\left[ \frac{i T_{1}}{2} (k_{31} - p_{31})^{2} + \frac{i T_{2}}{2} (k_{42} - p_{42})^{2} - i X_{12} \cdot P_{12} \right]} \\
	&\times \exp{\left[ -\frac{i T_{1}}{4} \left( \frac{t}{2} -m_{1}^{2} - m_{3}^{2} + 2 M_{1}^{2} \right) \right]} \\
	&\times \exp{\left[ -\frac{i T_{2}}{4} \left( \frac{t}{2} -m_{2}^{2} - m_{4}^{2} + 2 M_{2}^{2} \right) \right]}.
\end{split}
\end{equation}
This expression is Gaussian in the $k_{ij}$. The full integrand has the form ``Gaussian $\times$ function''. We resort to stationary methods to approximate the integral over the $k_{ij}$. The stationary point is
\begin{equation}
	\bar{k}_{31} = p_{31}, \qquad \bar{k}_{42} = p_{42}.
\end{equation}
At this stationary point, $\rho_{N}$ becomes a function of the (off-shell) external momenta $p_{ij}$,
\begin{equation}
	\rho_{N} = \frac{\mathcal{K}_{N}}{\sqrt{p_{31}^{2} p_{42}^{2} - (p_{31} \cdot p_{42})^{2}}}, \qquad \mathcal{K}_{N} = \left(p_{3 1} \cdots p_{3 1} \right) \cdot \nu_{N} \cdot \left(p_{4 2} \cdots p_{4 2}\right).
\end{equation}

So far, the un-integrated forward-JWKB scattering kernel looks like
\begin{equation}
\begin{split}
	\mathcal{S}_{T}(3,4|1,2) \approx \delta(P) \int \mathrm{d}X_{12} \, {}& \exp{\left[-i X_{12} \cdot P_{12} + \beta_{N} \rho_{N} \Gamma(\Delta - 1) \left( \frac{2}{B_{1 2}^{2}} \right)^{(\Delta - 1)} \right]} \\
	&\times \exp{\left[ -\frac{i T_{1}}{4} \left( \frac{t}{2} -m_{1}^{2} - m_{3}^{2} + 2 M_{1}^{2} \right) \right]} \\
	&\times \exp{\left[ -\frac{i T_{2}}{4} \left( \frac{t}{2} -m_{2}^{2} - m_{4}^{2} + 2 M_{2}^{2} \right) \right]}.
\end{split}
\label{STTX12}
\end{equation}
We defined $B_{12}$ as the part of $X_{12}$ that is orthogonal to any linear combination of the $x_{ij}$. But the effect of integration over the $x_{ij}$ was to replace $(x_{31}, x_{42})$ with $(T_{1} p_{31}, T_{2} p_{42})$. So now we have the decomposition
\begin{equation}
	X_{12} = B_{12} + T_{1} b_{31} p_{31} + T_{2} b_{42} p_{42}, \qquad B_{12} \cdot p_{31} = 0, \qquad B_{12} \cdot p_{42} = 0.
\end{equation}
The $X_{12}$ volume element becomes
\begin{equation}
	\mathrm{d}X_{12} = T_{1} T_{2} \sqrt{p_{31}^{2} p_{42}^{2} - (p_{31} \cdot p_{42})^{2}} \mathrm{d}B_{12} \mathrm{d}b_{31} \mathrm{d}b_{42};
\end{equation}
which we can write in terms of $\rho_{N}$,
\begin{equation}
	\mathrm{d}X_{12} = T_{1} T_{2} \left( \frac{\mathcal{K}_{N}}{\rho_{N}} \right) \mathrm{d}B_{12} \mathrm{d}b_{31} \mathrm{d}b_{42}.
\end{equation}
Note that
\begin{equation}
	X_{12} \cdot P_{12} = B_{12} \cdot P_{12} + T_{1} b_{31} (p_{31} \cdot P_{12}) + T_{2} b_{42} (p_{42} \cdot P_{12}).
\end{equation}
Since $\Sigma_{\text{two}}$ has no dependence on the $b_{ij}$, integration yields two one-dimensional Dirac deltas:
\begin{align}
	\int \mathrm{d}b_{31} \, \exp{\left[- i T_{1} b_{31} (p_{31} \cdot P_{12}) \right]} &= \frac{1}{T_{1}} \delta(p_{31} \cdot P_{12}); \\
	\int \mathrm{d}b_{42} \, \exp{\left[- i T_{2} b_{42} (p_{42} \cdot P_{12}) \right]} &= \frac{1}{T_{2}} \delta(p_{42} \cdot P_{12}).
\end{align}
We will examine later the constraints that these two Dirac deltas impose. Now the only part of the amplitude that depends on $(T_{1}, T_{2})$ are the second and third lines of (\ref{STTX12}). We must integrate over the moduli in order to obtain the integrated scattering kernel:
\begin{equation}
	\widehat{\mathcal{A}}(3, 4| 1, 2) \equiv \int\limits_{0}^{\infty} \mathrm{d}T_{1} \int\limits_{0}^{\infty} \mathrm{d}T_{2} \, \mathcal{S}_{T}(3, 4| 1, 2).
\end{equation}
Performing the integration over $(T_{1}, T_{2})$ yields
\begin{align}
	\int\limits_{0}^{\infty} \mathrm{d}T_{1} \, \exp{\left[ -\frac{i T_{1}}{4} \left( \frac{t}{2} - m_{1}^{2} - m_{3}^{2} + 2 M_{1}^{2}\right) \right]} &= \frac{8i}{t - 2m_{1}^{2} - 2m_{3}^{2} + 4 M_{1}^{2}}; \\
	\int\limits_{0}^{\infty} \mathrm{d}T_{2} \, \exp{\left[ -\frac{i T_{2}}{4} \left( \frac{t}{2} - m_{2}^{2} - m_{4}^{2} + 2 M_{2}^{2}\right) \right]} &= \frac{8i}{t - 2m_{2}^{2} - 2m_{4}^{2} + 4 M_{2}^{2}}.
\end{align}
The integral over $B_{12}$ remains:
\begin{equation}
	\widehat{\mathcal{A}} = \mathcal{N} \delta(P) \left( \frac{\mathcal{K}_{N}}{\rho_{N}} \right) \int \mathrm{d}B_{12} \, \exp{\left[-i B_{12} \cdot P_{12} + \beta_{N} \rho_{N} \Gamma(\Delta - 1) \left( \frac{2}{B_{1 2}^{2}} \right)^{(\Delta - 1)} \right]}; \label{AHatN}
\end{equation}
where we have collected some terms into an overall factor:
\begin{equation}
	\mathcal{N} \equiv \frac{(8i)^{2} \delta(p_{31} \cdot P_{12}) \delta(p_{42} \cdot P_{12})}{(t - 2m_{1}^{2} - 2m_{3}^{2} + 4 M_{1}^{2})(t - 2m_{2}^{2} - 2m_{4}^{2} + 4 M_{2}^{2})}.
\end{equation}
Before we put the external momenta on-shell, we need to truncate from $\widehat{\mathcal{A}}$ the part that is divergent on-shell.
\subsubsection{Truncation of External On-shell States}
In quantum field theory, truncation typically involves multiplying the scattering amplitude by a product of inverse propagators $(p_{j}^{2} + m_{j}^{2})$, and taking the limit $p_{j}^{2} \rightarrow - m_{j}^{2}$. This is done in order to remove the part that is divergent on-shell from the scattering amplitude. Since we have four external states, we need to remove four factors from $\widehat{\mathcal{A}}$.

From the relations
\begin{equation}
	p_{31} \cdot P_{12} = \frac{p_{3}^{2} - p_{1}^{2}}{2}, \qquad p_{42} \cdot P_{12} = \frac{p_{2}^{2} - p_{4}^{2}}{2};
\end{equation}
we see that the two Dirac deltas in $\mathcal{N}$ enforce the elasticity constraints
\begin{equation}
	p_{1}^{2} = p_{3}^{2}, \qquad p_{2}^{2} = p_{4}^{2}.
\end{equation}
Furthermore, on-shell we have
\begin{equation}
	p_{31}^{2} = \frac{t - 2 m_{1}^{2} - 2m_{3}^{2}}{4}, \qquad p_{42}^{2} = \frac{t - 2 m_{2}^{2} - 2m_{4}^{2}}{4};
\end{equation}
so then the denominators in $\mathcal{N}$ become
\begin{equation}
	\frac{8i}{t - 2m_{1}^{2} - 2m_{3}^{2} + 4 M_{1}^{2}} = \frac{2i}{p_{31}^{2} + M_{1}^{2}}, \qquad \frac{8i}{t - 2m_{2}^{2} - 2m_{4}^{2} + 4 M_{2}^{2}} = \frac{2i}{p_{42}^{2} + M_{2}^{2}}.
\end{equation}
These two factors have the form of (free) Feynman propagators for particles with momenta $(p_{31}, p_{42})$ and masses $(M_{1}, M_{2})$; they diverge when $p_{31}^{2} \rightarrow -M_{1}^{2}$ and $p_{42}^{2} \rightarrow -M_{2}^{2}$. We can think of these limits as ways to relate the external on-shell momenta to the ``internal'' masses $(M_{1}, M_{2})$:
\begin{equation}
	M_{1}^{2} = \frac{2m_{1}^{2} + 2m_{3}^{2} - t}{4}, \qquad M_{2}^{2} = \frac{2m_{2}^{2} + 2m_{4}^{2} - t}{4}.
\end{equation}
Note that these relations satisfy
\begin{equation}
	2M_{1}^{2} + 2M_{2}^{2} = m_{1}^{2} + m_{2}^{2} + m_{3}^{2} + m_{4}^{2} - t = s + u.
\end{equation}
After we enforce the elasticity constraints, we find
\begin{equation}
	M_{1}^{2} = m_{1}^{2} \left( 1 - \frac{t}{4 m_{1}^{2}} \right), \qquad M_{2}^{2} = m_{2}^{2} \left( 1 - \frac{t}{4 m_{2}^{2}} \right);
\end{equation}
and thus, in the forward-JWKB approximation (\ref{fJWKBLimit}) we have $M_{1} \approx m_{1}$ and $M_{2} \approx m_{2}$. The upshot of this discussion is that we can think of the factors in $\mathcal{N}$ as restricting the external momenta to be on-shell. Truncation is achieved by simply dropping $\mathcal{N}$ from (\ref{AHatN}). Thus, the \textit{truncated, on-shell, forward-JWKB scattering amplitude} $\mathcal{A}$ is given by
\begin{equation}
	\mathcal{A} = \delta(P) \left( \frac{\mathcal{K}_{N}}{\rho_{N}} \right) \int \mathrm{d}B_{12} \, \exp{\left[-i B_{12} \cdot P_{12} + \beta_{N} \rho_{N} \Gamma(\Delta - 1) \left( \frac{2}{B_{1 2}^{2}} \right)^{(\Delta - 1)} \right]}.
	\label{AHND}
\end{equation}
Recall that $B_{12}$ is a vector in $D$ dimensions subjected to two orthogonality constraints. Thus, the $B_{12}$ integral is over a $(D-2)$-dimensional volume. In \S\ref{sec4} and \S\ref{sec5} we evaluate this integral in $D = 3$ and $D = 4$.
\section{Forward-JWKB Amplitudes in $D = 3$\label{sec4}}
We now consider scattering in three spacetime dimensions. The coupling $g_{N}$ has units
\begin{equation}
	D = 3: \qquad [g_{N}] = \left( \frac{3 - 2N}{2} \right) [\text{mass}];
\end{equation}
and thus the coupling $\beta_{N}$ introduced in (\ref{betaN}) has units
\begin{equation}
	D = 3: \qquad [\beta_{N}] = 2[g_{N}] = \left(3 - 2N \right) [\text{mass}].
\end{equation}
For future reference we record the units of $\mathcal{K}_{N}$ and $\rho_{N}$ (in any number of dimensions):
\begin{equation}
	[\mathcal{K}_{N}] = 2N [\text{mass}], \qquad [\rho_{N}] = 2(N - 1) [\text{mass}].
\end{equation}
When $D = 3$ we have $\Delta = 1/2$, and thus (\ref{AHND}) takes the form
\begin{equation}
	\mathcal{A} = \beta_{N} \mathcal{K}_{N} \delta(P) \left( \frac{\sqrt{2 \pi}}{\sqrt{2 \pi} \beta_{N} \rho_{N}} \right) \int \mathrm{d}B_{1 2} \, \exp{\left[-i B_{12} \cdot P_{12} - \sqrt{2\pi} \beta_{N} \rho_{N} \vert B_{1 2} \vert \right]}.
\end{equation}
We recognize this as the Fourier transform of a massive, scalar propagator in one spacetime dimension along a space-like coordinate with ``mass'' given by $\sqrt{2\pi} \beta_{N} \rho_{N}$. The Fourier transform is just the familiar massive Feynman propagator:
\begin{equation}
	\mathcal{A}(s, t) = \delta(P) \left[ \frac{2 \beta_{N} \mathcal{K}_{N}(s)}{2 \pi \beta_{N}^{2} \rho_{N}^{2}(s) - t} \right] = \delta(P) \left[- \frac{2 \beta_{N} \mathcal{K}_{N}(s)}{t} \right] \left[ 1 - \frac{2 \pi \beta_{N}^{2} \rho_{N}^{2}(s)}{t} \right]^{-1}.
	\label{AHN3}
\end{equation}
In the second step we have extracted the expected tree-level massless singularity. Besides this singularity, the amplitude has a simple pole at $t = 2 \pi \beta_{N}^{2} \rho_{N}^{2}$. We now specialize to particular values of $N$ in order to study this singularity further.
\subsection{Exchange of Massless Scalar}
With $N = 0$, the matter particles exchange a massless scalar. The coupling $\beta_{0}$ has units
\begin{equation}
	[\beta_{0}] = 3 [\text{mass}].
\end{equation}
For the tensor in the kinetic operator, we have $\kappa_{0} = 1$, which leads to $\nu_{0} = 1$, and thus $\mathcal{K}_{0} = 1$. Hence,
\begin{equation}
	\rho_{0} = \frac{1}{\sqrt{p_{31}^{2} p_{42}^{2} - (p_{31} \cdot p_{42})^{2}}}.
\end{equation}
Using the on-shell identities
\begin{equation}
	p_{31}^{2} = -M_{1}^{2}, \qquad p_{42}^{2} = -M_{2}^{2}, \qquad p_{31} \cdot p_{42} = \frac{M_{1}^{2} + M_{2}^{2} - s}{2};
\end{equation}
leads to
\begin{equation}
	\rho_{0}(s) = \frac{2}{\sqrt{-\Lambda_{M}(s)}}, \qquad \Lambda_{M}(s) \equiv [s - (M_{1} - M_{2})^{2}] [s - (M_{1} + M_{2})^{2}].
\end{equation}
The singularity $t_{*} = 2 \pi \beta_{0}^{2} \rho_{0}^{2}(s_{*})$ leads to
\begin{equation}
	s_{*} = M_{1}^{2} + M_{2}^{2} + 2 M_{1} M_{2} \left(1 - \frac{2 \pi \beta_{0}^{2}}{M_{1}^{2} M_{2}^{2} t_{*}} \right)^{1/2}.
	\label{3s0}
\end{equation}
Using $u_{*} = 2m_{1}^{2} + 2m_{2}^{2} - t_{*} - s_{*}= 2M_{1}^{2} + 2M_{2}^{2} - s_{*}$ leads to
\begin{equation}
	u_{*} = M_{1}^{2} + M_{2}^{2} - 2 M_{1} M_{2} \left(1 - \frac{2 \pi \beta_{0}^{2}}{M_{1}^{2} M_{2}^{2} t_{*}} \right)^{1/2}.
\end{equation}
The product $s_{*} u_{*}$ gives
\begin{equation}
	s_{*} u_{*} = (M_{1} - M_{2})^{2}(M_{1} + M_{2})^{2} + \frac{8 \pi \beta_{0}^{2}}{t_{*}}
	= (m_{1} - m_{2})^{2}(m_{1} + m_{2})^{2} + \frac{8 \pi \beta_{0}^{2}}{t_{*}}.
	\label{3su0}
\end{equation}
Thus, if $t_{*} \leq 0$ then $s_{*}$ and $u_{*}$ are inside of the physical scattering region. However, continuation to $t_{*} > 0$ allows a window with real values of $s_{*}$ and $u_{*}$ as long as
\begin{equation}
	t_{*} > \frac{2 \pi \beta_{0}^{2}}{M_{1}^{2} M_{2}^{2}}.
\end{equation}
This lies outside of the physical scattering region and suggests a bound state.

In the forward-JWKB approximation (\ref{fJWKBLimit}), we have $M_{1} \approx m_{1}$ and $M_{2} \approx m_{2}$. When $t_{*} \leq 0$, we expect
\begin{equation}
	\frac{s_{*}}{m_{1} m_{2}} \text{ fixed}, \qquad \frac{u_{*}}{m_{1} m_{2}} \text{ fixed.}
\end{equation}
In order for this to hold, in $D = 3$ we must supplement (\ref{fJWKBLimit}) with
\begin{equation}
	\frac{\beta_{0}^{2}}{m_{1}^{2} m_{2}^{2} t_{*}} \text{ fixed} \quad \Longrightarrow \quad \frac{\beta_{0}^{2}}{m_{1}^{3} m_{2}^{3}} \rightarrow 0^{+};
\end{equation}
which suggest \textit{weak-coupling} in the $D = 3$ version of the forward-JWKB approximation.
\subsection{Exchange of Massless Vector}
With $N = 1$, the particles exchange a massless vector. The coupling $\beta_{1}$ now has units
\begin{equation}
	[\beta_{1}] = [\text{mass}].
\end{equation}
The tensor in the gauge-fixed kinetic operator is $(\kappa_{1})^{ab} = \eta^{ab}$, which leads to $(\nu_{1})_{ab} = \eta_{ab}$ and thus
\begin{equation}
	\mathcal{K}_{1} = p_{31} \cdot p_{42} = \frac{M_{1}^{2} + M_{2}^{2} - s}{2}.
\end{equation}
Hence,
\begin{equation}
	\rho_{1}(s) = Z_{1} Z_{2} \left[ \frac{M_{1}^{2} + M_{2}^{2} - s}{\sqrt{-\Lambda_{M}(s)}} \right];
\end{equation}
where we have included dimensionless charges $Z_{1}$ and $Z_{2}$ for each particle. The singularity $t_{*} = 2 \pi \beta_{1}^{2} \rho_{1}^{2}(s_{*})$ now leads to
\begin{equation}
	s_{*} = M_{1}^{2} + M_{2}^{2} + 2 M_{1} M_{2} \left(1 + \frac{2 \pi Z_{1}^{2} Z_{2}^{2} \beta_{1}^{2}}{t_{*}} \right)^{-1/2};
\end{equation}
and thus
\begin{equation}
	u_{*} = M_{1}^{2} + M_{2}^{2} - 2 M_{1} M_{2} \left(1 + \frac{2 \pi Z_{1}^{2} Z_{2}^{2} \beta_{1}^{2}}{t_{*}} \right)^{-1/2}.
\end{equation}
The product $s_{*} u_{*}$ now gives
\begin{align}
	s_{*}& u_{*} = (M_{1} - M_{2})^{2}(M_{1} + M_{2})^{2} + 4 M_{1}^{2} M_{2}^{2} \left( \frac{2 \pi Z_{1}^{2} Z_{2}^{2} \beta_{1}^{2}}{2 \pi Z_{1}^{2} Z_{2}^{2} \beta_{1}^{2} + t_{*}} \right) \nonumber \\
	&= (m_{1} - m_{2})^{2}(m_{1} + m_{2})^{2} + 4 m_{1}^{2} m_{2}^{2} \left(1 - \frac{t_{*}}{4 m_{1}^{2}} \right) \left(1 - \frac{t_{*}}{4 m_{2}^{2}} \right) \left( \frac{2 \pi Z_{1}^{2} Z_{2}^{2} \beta_{1}^{2}}{2 \pi Z_{1}^{2} Z_{2}^{2} \beta_{1}^{2} + t_{*}} \right).
\end{align}
Again, inside of the physical scattering region we have $t_{*} \leq 0$ and hence $s_{*}$ and $u_{*}$ are also inside of the physical scattering region. However, in order for $s_{*}$ and $u_{*}$ to be real and finite, we must require
\begin{equation}
	{-t_{*}} > 2 \pi Z_{1}^{2} Z_{2}^{2} \beta_{1}^{2}.
\end{equation}
If we analytically continue to $t_{*} > 0$, we find $s_{*}$ and $u_{*}$ real for any value of $t_{*}$.

With the massless vector exchange, in $D = 3$ we must supplement (\ref{fJWKBLimit}) with
\begin{equation}
	\frac{\beta_{1}^{2}}{t_{*}} \text{ fixed} \quad \Longrightarrow \quad \frac{\beta_{1}^{2}}{m_{1} m_{2}} \rightarrow 0^{+};
\end{equation}
which also suggest \textit{weak-coupling}.
\subsection{Exchange of Massless Symmetric Tensor}
A massless symmetric (traceless) tensor has no propagating degrees of freedom in $D = 3$. Setting $N = 2$ corresponds to a sort of analytic continuation. Whatever the result gives, it should not have an interpretation in terms of propagating gravitons. The coupling $\beta_{2}$ has units
\begin{equation}
	[\beta_{2}] = - [\text{mass}].
\end{equation}
When $N = 2$, the tensor in the gauge-fixed kinetic operator is
\begin{equation}
	(\kappa_{2})^{a_{1}b_{1} a_{2} b_{2}} = \frac{1}{2} \left( \eta^{a_{1} a_{2}} \eta^{b_{1} b_{2}} + \eta^{a_{1}b_{2}} \eta^{b_{1}a_{2}} - \eta^{a_{1}b_{1}} \eta^{a_{2}b_{2}} \right).
\end{equation}
In $D = 3$, we find
\begin{equation}
	D = 3: \qquad (\nu_{2})_{a_{1}b_{1} a_{2} b_{2}} = \frac{1}{2} \left( \eta_{a_{1} a_{2}} \eta_{b_{1} b_{2}} + \eta_{a_{1}b_{2}} \eta_{b_{1}a_{2}} - 2\eta_{a_{1}b_{1}} \eta_{a_{2}b_{2}} \right).
\end{equation}
Thus, in $D = 3$ we have
\begin{equation}
	D = 3: \qquad \mathcal{K}_{2} = (p_{31} \cdot p_{42})^{2} - p_{31}^{2} p_{42}^{2} = \frac{1}{4} \Lambda_{M}(s);
	\label{D3K2}
\end{equation}
and hence
\begin{equation}
	D = 3: \qquad \rho_{2}(s) = -\frac{1}{2} \sqrt{-\Lambda_{M}(s)}.
	\label{D3rho2}
\end{equation}
The singularity $t_{*} = 2 \pi \beta_{2}^{2} \rho_{2}^{2}(s_{*})$ leads to
\begin{equation}
	s_{*} = M_{1}^{2} + M_{2}^{2} + 2 M_{1} M_{2} \left( 1 - \frac{t_{*}}{2\pi M_{1}^{2} M_{2}^{2} \beta_{2}^{2}} \right)^{1/2};
	\label{3s2}
\end{equation}
and thus
\begin{equation}
	u_{*} = M_{1}^{2} + M_{2}^{2} - 2 M_{1} M_{2} \left( 1 - \frac{t_{*}}{2\pi M_{1}^{2} M_{2}^{2} \beta_{2}^{2}} \right)^{1/2}.
	\label{3u2}
\end{equation}
We look at the product $s_{*} u_{*}$:
\begin{equation}
	s_{*} u_{*} = (M_{1} - M_{2})^{2}(M_{1} + M_{2})^{2} + \frac{2 t_{*}}{\pi \beta_{2}^{2}}
	= (m_{1} - m_{2})^{2}(m_{1} + m_{2})^{2} + \frac{2 t_{*}}{\pi \beta_{2}^{2}}.
	\label{3su2}
\end{equation}
Note the similarity between (\ref{3su0}) and (\ref{3su2}). Indeed, under the replacement
\begin{equation}
	\frac{2\pi \beta_{0}^{2}}{t_{*}} \longleftrightarrow \frac{t_{*}}{2\pi \beta_{2}^{2}};
\end{equation}
we find complete agreement. It follows that if $t_{*} \leq 0$, then $s_{*}$ and $u_{*}$ are inside of the physical scattering region. If we let $t_{*} > 0$, then we must have
\begin{equation}
	t_{*} \leq 2 \pi M_{1}^{2} M_{2}^{2} \beta_{2}^{2};
\end{equation}
in order for $s_{*}$ and $u_{*}$ to be real.

Looking at (\ref{3s2}) and enforcing the forward-JWKB approximation (\ref{fJWKBLimit}), it is easy to conclude that $s_{*} \approx (m_{1} + m_{2})^{2}$. This assumes that $m_{1} m_{2} \beta_{2}^{2}$ is kept fixed in the forward-JWKB approximation. A more correct statement is that $s_{*} / (m_{1} m_{2})$ is kept fixed, which leads to
\begin{equation}
	\frac{t_{*}}{m_{1}^{2} m_{2}^{2} \beta_{2}^{2}} \text{ fixed} \quad \Longrightarrow \quad m_{1} m_{2} \beta_{2}^{2} \rightarrow 0^{+};
\end{equation}
which, yet again, suggests \textit{weak-coupling}. However, since the coupling $\beta_{2}$ appears in denominators in (\ref{3s2}) and (\ref{3su2}), it seems that this weak-coupling phenomenon does not arise from perturbative contributions involving the exchange of propagating quanta.
\section{Forward-JWKB Amplitudes in $D = 4$\label{sec5}}
In $D = 3$ we found that the forward-JWKB scattering amplitude has one extra singularity. As we will see shortly, in $D = 4$ we have an infinite number of singularities. For $N = 0$ and $N = 1$ the structure of these singularities in $D = 4$ is related to the singularity in $D = 3$. When $N = 2$ we find an infinite number of singularities that have a very different interpretation from the singularity in $D = 3$.

But as soon as we set $D = 4$ (i.e. $\Delta = 1$), we find trouble inside of the exponential in (\ref{AHND}). To get around this issue, we work instead in $D = 4 + 2 \epsilon$ with $\epsilon > 0$. The coupling parameter $g_{N}$ and $\beta_{N}$ have units
\begin{equation}
	D = 4 + 2 \epsilon: \qquad [g_{N}] = \left( 1 - \epsilon - N \right) [\text{mass}], \qquad [\beta_{N}] = 2 \left( 1 - \epsilon - N \right) [\text{mass}].
\end{equation}
It is convenient to extract from $\beta_{N}$ the four-dimensional coupling $\alpha_{N}$ by introducing a constant $\mu$ with units of mass:
\begin{equation}
	\beta_{N} = \alpha_{N} \left( \frac{1}{\mu^{2}} \right)^{\epsilon}.
\end{equation}
Inside the exponential in (\ref{AHND}), we have
\begin{align}
	\beta_{N} \Gamma(\Delta - 1) \left( \frac{2}{B_{12}^{2}} \right)^{(\Delta - 1)} &= \alpha_{N} \Gamma(\epsilon) \left( \frac{2}{\mu^{2} B_{12}^{2}} \right)^{\epsilon} \nonumber \\
	&\approx \alpha_{N} \Gamma(\epsilon) + \alpha_{N} \log{\left( \frac{2}{\mu^{2} B_{12}^{2}} \right)} + O(\epsilon);
\end{align}
where in the second line we have expanded near $\epsilon = 0$ and kept the leading logarithm. The amplitude near four-dimensions becomes
\begin{equation}
	\mathcal{A}(s, t) \approx \delta(P) \left[ \frac{\mathcal{K}_{N}}{\rho_{N}} \right] \exp{\left[ \alpha_{N} \rho_{N} \Gamma(\epsilon) \right]} \int \mathrm{d}B_{12} \left( \frac{2}{\mu^{2} B_{12}^{2}} \right)^{\alpha_{N} \rho_{N}} \exp{\left( - i B_{12} \cdot P_{12} \right)}.
\end{equation}
Note that the divergent part has factored out and appears inside an exponential. The integral over $B_{12}$ is now over a $D - 2 \approx 2$ dimensional volume. Integration yields
\begin{equation}
	\mathcal{A}(s, t) \approx \delta(P) \left[ \frac{\alpha_{N} \mathcal{K}_{N}(s)}{\mu^{2}} \right] \exp{\left[ \alpha_{N} \rho_{N}(s) \Gamma(\epsilon) \right]} \frac{\Gamma[1 - \alpha_{N} \rho_{N}(s)]}{\Gamma[1 + \alpha_{N} \rho_{N}(s)]} \left( - \frac{t}{2 \mu^{2}} \right)^{(\alpha_{N} \rho_{N}(s) - 1)}.
	\label{AHN4notree}
\end{equation}
A convenient way to write this result is
\begin{equation}
	\mathcal{A}(s, t) \approx \mathcal{A}_{\text{tree}}(s, t) \exp{\left[ \alpha_{N} \rho_{N}(s) \Gamma(\epsilon) \right]} \frac{\Gamma[1 - \alpha_{N} \rho_{N}(s)]}{\Gamma[1 + \alpha_{N} \rho_{N}(s)]} \left( - \frac{t}{2 \mu^{2}} \right)^{\alpha_{N} \rho_{N}(s)};
	\label{AHN4}
\end{equation}
where we have collected the tree-level contribution into an overall factor,
\begin{equation}
	\mathcal{A}_{\text{tree}}(s, t) = \delta(P) \left[ - \frac{2 \alpha_{N} \mathcal{K}_{N}(s)}{t} \right];
\end{equation}
which exhibits the familiar massless pole at $t = 0$. Indeed, the result (\ref{AHN4}) exhibits an infinite number of singularities from the poles of the Euler Gamma function in the numerator. In order to make all of the singularities manifest, it is useful to decompose the amplitude (\ref{AHN4notree}) into partial waves:
\begin{equation}
	\mathcal{A}(s, t) = \sum_{l = 0}^{\infty} (2l + 1) \mathcal{A}_{l}(s) P_{l}(z_{s}), \qquad z_{s} = \cos{(\theta_{s})};
\end{equation}
where $P_{l}$ is a Legendre polynomial, and the partial amplitude $\mathcal{A}_{l}$ is given by
\begin{equation}
	\mathcal{A}_{l}(s) = \frac{1}{2} \int\limits_{-1}^{1} \mathrm{d}z_{s} \, \mathcal{A}(s, t) P_{l}(z_{s}).
\end{equation}
Using
\begin{equation}
	{-t} = \frac{\Lambda_{12}(s)}{s} \left( \frac{1 - z_{s}}{2} \right);
\end{equation}
and
\begin{equation}
	P_{l}(z_{s}) = \sum_{k = 0}^{l} \frac{\Gamma(1 + l)}{\Gamma(1 + k) \Gamma(1 + l - k)} \frac{\Gamma(-l)}{\Gamma(1 + k) \Gamma(-l-k)} \left( \frac{1 - z_{s}}{2} \right)^{k};
\end{equation}
leads to the partial amplitude
\begin{equation}
	\mathcal{A}_{l}(s) = \delta(P) \frac{\sqrt{- \Lambda_{M}(s)}}{2 \mu^{2}} \exp{\left[ \alpha_{N} \rho_{N}(s) \Gamma(\epsilon) \right]} \left[ \frac{\Lambda_{12}(s)}{2 \mu^{2} s} \right]^{(\alpha_{N} \rho_{N} - 1)} \frac{\Gamma[1 - \alpha_{N} \rho_{N}(s) + l]}{\Gamma[1 + \alpha_{N} \rho_{N}(s) + l]}.
\end{equation}
This partial amplitude has a singularity whenever
\begin{equation}
	1 - \alpha_{N} \rho_{N}(s_{nl}) + l = - n, \qquad n = 0, 1, 2, \ldots \qquad l = 0, 1, 2, \ldots
\end{equation}
At first glance, we see a close similarity between this singularity condition in $D = 4$ and the condition $t = 2 \pi \beta_{N}^{2} \rho_{N}^{2}(s_{*})$ in $D = 3$. We now consider specific values of $N$.
\subsection{Exchange of Massless Scalar}
A massless scalar field has the same dynamics in $D = 3$ and $D = 4$. Thus, we again have
\begin{equation}
	\rho_{0}(s) = \frac{2}{\sqrt{-\Lambda_{M}(s)}}, \qquad \Lambda_{M}(s) \equiv [s - (M_{1} - M_{2})^{2}] [s - (M_{1} + M_{2})^{2}].
	\label{rho0D4}
\end{equation}
In $D = 4$, the coupling $\alpha_{0}$ has units
\begin{equation}
	[\alpha_{0}] = 2 [\text{mass}].
\end{equation}
The singularity condition becomes $1 - \alpha_{0} \rho_{0}(s_{nl}) + l = -n$. Comparing this to $t_{*} = 2 \pi \beta_{0}^{2} \rho_{0}^{2}(s_{*})$ in $D = 3$, we can find the solution for $s_{nl}$ by using (\ref{3s0}) with the replacement
\begin{equation}
	\frac{2 \pi \beta_{0}^{2}}{t_{*}} \longrightarrow \frac{\alpha_{0}^{2}}{(n + l + 1)^{2}}.
\end{equation}
Note that this replacement is only valid when $t_{*} > 0$ (i.e. outside of the physical scattering region), since the right-hand side is always positive. Thus, in $D = 4$ we have the infinite sequence
\begin{equation}
	s_{nl} = M_{1}^{2} + M_{2}^{2} + 2 M_{1} M_{2} \left(1 - \frac{\alpha_{0}^{2}}{M_{1}^{2} M_{2}^{2} (n + l + 1)^{2}} \right)^{1/2}.
	\label{sJ0}
\end{equation}
Using $s_{nl} + u_{nl} = 2M_{1}^{2} + 2M_{2}^{2}$ leads to
\begin{equation}
	u_{nl} = M_{1}^{2} + M_{2}^{2} - 2 M_{1} M_{2} \left(1 - \frac{\alpha_{0}^{2}}{M_{1}^{2} M_{2}^{2} (n + l + 1)^{2}} \right)^{1/2}.
\end{equation}
Unlike in $D = 3$, the infinite sequences $(s_{nl}, u_{nl})$ always lie outside of the physical scattering region. This suggest an interpretation as bound state singularities.

The singularities $(s_{nl}, u_{nl})$ lie outside of the physical scattering region, but we can still look for the requirement that $s_{nl} / (m_{1} m_{2})$ is fixed in the forward-JWKB approximation (\ref{fJWKBLimit}):
\begin{equation}
	\frac{s_{nl}}{m_{1} m_{2}} \text{ fixed} \quad \Longrightarrow \quad \frac{\alpha_{0}^{2}}{M_{1}^{2} M_{2}^{2} (n + l + 1)^{2}} \text{ fixed}.
\end{equation}
Since the semiclassical approximation involves large quantum numbers, we have
\begin{equation}
	n + l \rightarrow \infty \quad \Longrightarrow \quad \frac{\alpha_{0}}{M_{1} M_{2}} \rightarrow \infty;
\end{equation}
which suggests \textit{strong-coupling} in the forward-JWKB regime in $D = 4$.
\subsection{Exchange of Massless Vector}
A massless vector field has two physical polarizations in $D = 4$, and one in $D = 3$. However, the gauge-fixed kinetic operators are the same in any number of dimensions. Again, we have
\begin{equation}
	\rho_{1}(s) = Z_{1} Z_{2} \left[ \frac{M_{1}^{2} + M_{2}^{2} - s}{\sqrt{-\Lambda_{M}(s)}} \right];
\end{equation}
where, again, we have introduced dimensionless charges $Z_{1}$ and $Z_{2}$. The singularity condition becomes $1 - \alpha_{1} \rho_{1}(s_{nl}) + l = -n$, which is again analogous to the singularity condition in $D = 3$ with the replacement
\begin{equation}
	\frac{2 \pi \beta_{1}^{2}}{t_{*}} \longrightarrow \frac{\alpha_{1}^{2}}{(n + l + 1)^{2}}.
\end{equation}
Hence, in $D = 4$ we have
\begin{equation}
	s_{nl} = M_{1}^{2} + M_{2}^{2} + 2 M_{1} M_{2} \left(1 + \frac{Z_{1}^{2} Z_{2}^{2} \alpha_{1}^{2}}{(n + l + 1)^{2}} \right)^{-1/2}.
	\label{sJ1}
\end{equation}
We find that $s_{nl}$ is always outside of the physical scattering region.

Keeping $s_{nl}/(m_{1} m_{2})$ fixed in the forward-JWKB approximation requires
\begin{equation}
	\frac{\alpha_{1}^{2}}{(n + l + 1)^{2}} \text{ fixed}.
\end{equation}
If $n + l \rightarrow \infty$, we must also have $\alpha_{1} \rightarrow \infty$. Thus, in $D = 4$ we again find \textit{strong-coupling}.
\subsection{Exchange of Massless Symmetric Tensor}
A massless symmetric (traceless) tensor has two physical polarizations in $D = 4$. The kinetic operator (and thus the tensor $\kappa_{2}$) are the same as in $D = 3$. However, in $D = 4$, the $\nu_{2}$ tensor gives
\begin{equation}
	D = 4: \qquad (\nu_{2})_{a_{1}b_{1} a_{2} b_{2}} = \frac{1}{2} \left( \eta_{a_{1} a_{2}} \eta_{b_{1} b_{2}} + \eta_{a_{1}b_{2}} \eta_{b_{1}a_{2}} - \eta_{a_{1}b_{1}} \eta_{a_{2}b_{2}} \right).
\end{equation}
This leads to a different $\mathcal{K}_{2}$ from (\ref{D3K2}),
\begin{equation}
	D = 4: \qquad \mathcal{K}_{2} = (p_{31} \cdot p_{42})^{2} - \frac{1}{2} p_{31}^{2} p_{42}^{2} = \frac{1}{4} \left[ (M_{1}^{2} + M_{2}^{2} - s)^{2} - 2 M_{1}^{2} M_{2}^{2} \right];
\end{equation}
and thus, a different $\rho_{2}(s)$ from (\ref{D3rho2}),
\begin{equation}
	D = 4: \qquad \rho_{2}(s) = \frac{1}{2} \left[ \frac{(M_{1}^{2} + M_{2}^{2} - s)^{2} - 2 M_{1}^{2} M_{2}^{2}}{\sqrt{-\Lambda_{M}(s)}} \right].
\end{equation}
In $D = 4$ the coupling $\alpha_{2}$ has units
\begin{equation}
	[\alpha_{2}] = -2 [\text{mass}].
\end{equation}
Solving the singularity condition $1 - \alpha_{2} \rho_{2}(s_{nl}) + l = -n$ leads to
\begin{equation}
	s_{nl} = M_{1}^{2} + M_{2}^{2} + 2M_{1}M_{2} \left[ \frac{1}{2} + \left(1 + \sqrt{1 + \frac{2 M_{1}^{2} M_{2}^{2} \alpha_{2}^{2}}{(n + l + 1)^{2}}} \right)^{-1} \right]^{1/2}.
	\label{sJ2}
\end{equation}
Again, we find that the infinite sequence of singularities $s_{nl}$ lie outside of the physical scattering region.

Keeping $s_{nl}/(m_{1}m_{2})$ fixed in the forward-JWKB approximation requires
\begin{equation}
	\frac{M_{1}^{2} M_{2}^{2} \alpha_{2}^{2}}{(n + l + 1)^{2}} \text{ fixed}.
\end{equation}
If $n + l \rightarrow \infty$, then also $M_{1} M_{2} \alpha_{2} \rightarrow \infty$, which yet again suggest \textit{strong-coupling} in $D = 4$. Moreover, for spin 2 interactions the product $m_{i} \alpha_{2}$ corresponds to the Schwarzschild radius $r_{i}$ of a particle with mass $m_{i}$. The product $M_{1} M_{2} \alpha_{2} \approx m_{1} m_{2} \alpha_{2}$ can be interpreted as the ratio of the Schwarzschild radius of one particle to the Compton wavelength of the other:
\begin{equation}
	m_{1} m_{2} \alpha_{2} = \frac{r_{1}}{\lambda_{2}} = \frac{r_{2}}{\lambda_{1}}.
\end{equation}
Thus, $m_{1} m_{2} \alpha_{2} \rightarrow \infty$ is also the regime of large Schwarzschild radii.
\section{Exchange of Heavy Scalar\label{sec6}}
In the previous two sections we studied a system with two non-identical heavy scalar particles interacting via the exchange of massless quanta with spin $0$, $1$ or $2$. For these three cases, the truncated forward-JWKB scattering amplitude has the same general form (see (\ref{AHN3}) and (\ref{AHN4})). We now consider a system with two non-identical heavy scalar particles that exchange heavy scalar quanta.

We can repeat most of the steps as before to derive the path integral analogous to (\ref{FSeff}). The tensor $\kappa_{0}$ for a massive scalar is the same as for a massless scalar, so again we have $\kappa_{0} = 1$, $\nu_{0} = 1$ and thus $\mathcal{K}_{0} = 1$. Thus, the kinetic term for the mediating field $h$ is
\begin{equation}
	S_{\text{kin}}[h] = \frac{1}{2} \int \int \mathrm{d}x \mathrm{d}y \, h(x) K_{M}(x|y) h(y), \qquad K_{M}(x|y) \equiv \delta(x - y) \left(- \frac{1}{2} \partial^{2} + \frac{1}{2} M^{2} \right).
\end{equation}
Here $M$ is a constant with units of mass. The functional integral over $h$ gives rise to interaction terms analogous to (\ref{S10}) and (\ref{S20}). Just as before, we will \textbf{ignore the contributions from the self-interactions}. The two-body interaction term is
\begin{equation}
	S_{2}[q_{1}, q_{2}] = g_{0}^{2} \int \int \mathrm{d}\tau_{1} \mathrm{d}\tau_{2} \, G_{M}[q_{1}(\tau_{1}) | q_{2}(\tau_{2})];
\end{equation}
where $G_{M} = (K_{M})^{-1}$ is the free massive scalar Green function.

In the forward-JWKB approximation, we evaluate $S_{\text{two}}$ at the forward paths (\ref{fpaths}):
\begin{equation}
	\Sigma_{\text{two}} = S_{\text{two}}[f_{1}, f_{2}] = g_{0}^{2} \int \int \mathrm{d}\tau_{1} \mathrm{d}\tau_{2} \, G_{M}[f_{1}(\tau_{1}) | f_{2}(\tau_{2})].
\end{equation}
Recall the expression for $G_{M}$ in terms of a Schwinger integral:
\begin{equation}
	G_{M}(x|y) = M^{(D-2)} \int\limits_{0}^{\infty} \mathrm{d}T \left( - \frac{i}{T} \right)^{D/2} \exp{\left[ \frac{i}{2 T} M^{2}(y - x)^{2} - \frac{i T}{2} \right]}.
\end{equation}
Since the separation of the particles is very large in the forward-JWKB approximation, we can still integrate over $(\tau_{1}, \tau_{2})$ with stationary methods. The result gives
\begin{equation}
	\Sigma_{\text{two}} \approx 2 \pi g_{0}^{2} \left[ \frac{T_{1} T_{2}}{\sqrt{x_{31}^{2} x_{42}^{2} - (x_{31} \cdot x_{42})^{2}}} \right] M^{(D - 4)} \int\limits_{0}^{\infty} \mathrm{d}T \left( - \frac{i}{T} \right)^{\Delta} \exp{\left[ \frac{i}{2 T} M^{2}B_{12}^{2} - \frac{i T}{2} \right]}.
\end{equation}
with $\Delta = (D - 2) / 2$. In the regime $M^{2} B_{12}^{2} \rightarrow \infty$ (heavy messenger and large separations), the integral over $T$ is also done with stationary methods:
\begin{equation}
	\Sigma_{\text{two}} \approx -i \sqrt{2 \pi} \beta_{0} \rho_{0} M^{(D - 4)} \left( i M \sqrt{-B_{12}^{2}} \right)^{(3 - D)/2} \exp{\left( -i M \sqrt{-B_{12}^{2}}\right)};
	\label{Sigma2Massive}
\end{equation}
where we have used
\begin{equation}
	\beta_{0} = 2 \pi g_{0}^{2}, \qquad \rho_{0} = \frac{T_{1} T_{2}}{\sqrt{x_{31}^{2} x_{42}^{2} - (x_{31} \cdot x_{42})^{2}}}.
\end{equation}
One can recognize $\Sigma_{\text{two}}$ in (\ref{Sigma2Massive}) as being proportional to a long-distance massive propagator in $D-2$ dimensions (i.e. given by the asymptotic expansion of the familiar Bessel function). The truncated on-shell forward-JWKB scattering amplitude gives
\begin{equation}
	\mathcal{A} = \delta(P) \left( \frac{1}{\rho_{0}} \right) \int \mathrm{d}B_{1 2} \, \exp{\left( - i B_{1 2} \cdot P_{1 2} \right)} \left[-1 + \exp{\left(i \Sigma_{\text{two}}\right)} \right];
	\label{DAvarphi}
\end{equation}
with $\Sigma_{\text{two}}$ given by (\ref{Sigma2Massive}) and we have subtracted the disconnected part. Note that unlike the massless exchange, with the massive exchange $\Sigma_{\text{two}}$ is not explicitly divergent when $D = 4$.
\subsection{Three Spacetime Dimensions}
When $D = 3$, we have
\begin{equation}
	\Sigma_{\text{two}} \approx -i \left( \frac{\sqrt{2 \pi} \beta_{0} \rho_{0}}{M} \right) \exp{\left( -i M \sqrt{-B_{12}^{2}}\right)}.
\end{equation}
So then
\begin{equation}
	{-1} + \exp{(i \Sigma_{\text{two}})} = \sum_{L = 0}^{\infty} \frac{1}{L! (L+1)} \left( \frac{\sqrt{2 \pi} \beta_{0} \rho_{0}}{M} \right)^{(L+1)} \exp{\left( -i (L+1) M \sqrt{-B_{12}^{2}}\right)};
\end{equation}
which we recognize as the sum of one-dimensional massive scalar propagators with mass $(L+1)M$. Thus, after integration over $B_{12}$ we find
\begin{equation}
	\mathcal{A}(s, t) = \beta_{0} \delta(P) \sum_{L = 0}^{\infty} \frac{1}{L!} \left[ \frac{\sqrt{2 \pi} \beta_{0} \rho_{0}(s)}{M} \right]^{L} \left[ \frac{2}{(L+1)^{2} M^{2} - t} \right].
	\label{3Avarphi}
\end{equation}
This result has an infinite number of singularities given by $t = (L+1)^{2}M^{2}$, which can be recognized as the branch points of the $(L+1)$-mass branch cuts. We also have the singularities whenever $\rho_{0}(s_{*}) \rightarrow \infty$, which correspond to $s_{*} = (M_{1} \pm M_{2})^{2}$. It is interesting that instead of getting a whole branch cut (a continuum of singularities), in the forward-JWKB approximation we seem to only get the branch point (a single singularity).

Using the relation
\begin{equation}
	\frac{2}{(L+1)^{2} M^{2} - t} = \frac{1}{\sqrt{t}} \left[ \frac{1}{(L+1)M - \sqrt{t}} - \frac{1}{(L+1)M + \sqrt{t}} \right];
\end{equation}
and the incomplete Euler Gamma function,
\begin{equation}
	\Gamma_{\text{inc}}(z, a) = \int_{0}^{a} \mathrm{d}T \left( \frac{1}{T} \right)^{1-z} \exp{(-T)} = a^{z} \sum_{n = 0}^{\infty} \frac{1}{n!} \frac{(-a)^{n}}{(n + z)};
\end{equation}
we can rewrite (\ref{3Avarphi}) as
\begin{equation}
	\mathcal{A}(s, t) = \delta(P) \left( \frac{\beta_{0}}{M \sqrt{t}} \right) \left[ \mathcal{R}_{+}(s, t) + \mathcal{R}_{-}(s, t) \right];
\end{equation}
where
\begin{equation}
	\mathcal{R}_{\pm}(s, t) \equiv \pm \left[ - \frac{\sqrt{2 \pi} \beta_{0} \rho_{0}(s)}{M} \right]^{R_{\pm}(t)} \Gamma_{\text{inc}}\left[ -R_{\pm}(t),  - \frac{\sqrt{2 \pi} \beta_{0} \rho_{0}(s)}{M} \right];
	\label{615}
\end{equation}
with
\begin{equation}
	R_{\pm}(t) \equiv -1 \pm \sqrt{\frac{t}{M^{2}}}.
\end{equation}
This form of the amplitude is akin to Regge behavior, with the Regge poles being the multi-mass branch points.
\subsection{Five Spacetime Dimensions}
We can set $D = 5$ in (\ref{Sigma2Massive}) and find no explicit divergences. However, as a precaution we work in $D = 5 - 4 \varepsilon$ with $\varepsilon > 0$. Just like we did before in $D = 4 + 2 \epsilon$, we extract the $D = 5$ coupling $\gamma_{0}$ by introducing a constant $\mu$ with units of mass:
\begin{equation}
	\beta_{0} = \gamma_{0} \mu^{4 \varepsilon}.
\end{equation}
Note that $\gamma_{0}$ has units of mass. Then, $\Sigma_{\text{two}}$ becomes
\begin{equation}
	\Sigma_{\text{two}} \approx -i \sqrt{2 \pi} M\gamma_{0} \rho_{0} \left( \frac{\mu}{M} \right)^{4\varepsilon} \left(i M \sqrt{-B_{12}^{2}}\right)^{(2\varepsilon - 1)} \exp{\left(- i M \sqrt{-B_{12}^{2}}\right)};
\end{equation}
and thus
\begin{equation}
\begin{split}
	{-1} + \exp{(i \Sigma_{\text{two}})} = \sum_{L = 0}^{\infty} \frac{1}{\Gamma(L + 2)} {}& \left[ \sqrt{2 \pi} M \gamma_{0} \rho_{0} \left( \frac{\mu}{M} \right)^{4\varepsilon} \right]^{(L+1)} \left(i M \sqrt{-B_{12}^{2}}\right)^{(L+1)(2\varepsilon - 1)} \\
	&\times \exp{\left(- i (L+1) M \sqrt{-B_{12}^{2}}\right)}.
\end{split}
\end{equation}
In order to perform the $B_{12}$ integral in (\ref{DAvarphi}), we write the exponential in the second line as an infinite sum too. This leads to a double sum involving powers of $B_{12}$:
\begin{equation}
\begin{split}
	{-1} + \exp{(i \Sigma_{\text{two}})} = \sum_{L = 0}^{\infty} \sum_{n = 0}^{\infty} {}& \frac{(-1)^{n}(L+1)^{(n - 1)}}{\Gamma(L + 1)\Gamma(n+1)} \left[ \sqrt{2 \pi} M \gamma_{0} \rho_{0} \left( \frac{\mu}{M} \right)^{4\varepsilon} \right]^{(L+1)} \\
	&\times \left(\frac{1}{2}\right)^{\theta_{nL}} \left( \frac{2}{M^{2} B_{12}^{2}} \right)^{\theta_{nL}};
\end{split}
\end{equation}
with
\begin{equation}
	\theta_{nL} \equiv \frac{(L+1)(1 - 2\varepsilon)}{2} - \frac{n}{2}.
\end{equation}
Taking the Fourier transform of each power yields
\begin{equation}
\begin{split}
	\mathcal{A} = \frac{1}{\rho_{0}} \left( \frac{1}{M^{2}} \right)^{(3 - 4\varepsilon)/2} \delta(P) \sum_{L = 0}^{\infty} \sum_{n = 0}^{\infty} {}& \frac{(-1)^{n}(L+1)^{(n - 1)}}{\Gamma(L + 1) \Gamma(n+1)} \left[ \sqrt{2 \pi} M \gamma_{0} \rho_{0} \left( \frac{\mu}{M} \right)^{4\varepsilon} \right]^{(L+1)} \\
	&\times \left(\frac{1}{2}\right)^{\theta_{nL}} \frac{\Gamma(\omega_{nL})}{\Gamma(\theta_{nL})} \left( - \frac{2M^{2}}{t} \right)^{\omega_{nL}};
\end{split}
\end{equation}
with
\begin{equation}
	\omega_{nL} \equiv \frac{3}{2} - 2\varepsilon - \theta_{nL}.
\end{equation}
From this result, we anticipate a divergence whenever $\omega_{nL} = -l$ with $l = 0,1,2, \ldots$ or
\begin{equation}
	L = \frac{2 - 2\varepsilon + 2 l + n}{1 - 2\varepsilon}, \qquad l = 0, 1, 2, \ldots \qquad n = 0, 1, 2, \ldots
\end{equation}
Hence, if we take $\varepsilon \rightarrow 0$ we expect to find a divergence when $L \geq 2 $. To explicitly see this, we keep $L$ fixed and split the sum over $n$ into even and odd parts:
\begin{equation}
	\mathcal{A} = \frac{1}{M^{3}\rho_{0}} M^{4\varepsilon} \delta(P) \sum_{L = 0}^{\infty} \frac{1}{\Gamma(L + 1)} \left[ \sqrt{2 \pi} M \gamma_{0} \rho_{0} \left( \frac{\mu}{M} \right)^{4\varepsilon} \right]^{(L+1)} [\mathcal{E}_{L}(t) + \mathcal{O}_{L}(t)].
\end{equation}
We find
\begin{align}
	\mathcal{E}_{L}(t) = {}& \frac{2^{(2 \varepsilon - 1)(L + 1)/2}}{(L + 1)} \frac{\Gamma\left( \frac{2 - 2\varepsilon - (1 - 2\varepsilon) L}{2} \right)}{\Gamma\left( \frac{(L + 1)(1 - 2\varepsilon)}{2} \right)} \left( - \frac{2M^{2}}{t} \right)^{(2 - 2\varepsilon - (1 - 2\varepsilon)L)/2} \nonumber \\
	&\times {}_{2}F_{1}\left( \frac{2 - 2\varepsilon - (1 - 2\varepsilon)L}{2}, \frac{1 + 2\varepsilon - (1 - 2\varepsilon)L}{2}; \frac{1}{2}; \frac{(L+1)^{2}M^{2}}{t} \right); \\
	\mathcal{O}_{L}(t) = {}& - 2^{[(2 \varepsilon - 1)(L + 1) + 1]/2} \frac{\Gamma\left( \frac{3 - 2\varepsilon - (1 - 2\varepsilon)L}{2} \right)}{\Gamma \left( \frac{L(1 - 2\varepsilon) - 2\varepsilon}{2} \right)} \left( - \frac{2M^{2}}{t} \right)^{(3 - 2\varepsilon - (1 - 2\varepsilon)L)/2} \nonumber \\
	&\times {}_{2}F_{1}\left( \frac{3 - 2\varepsilon - (1 - 2\varepsilon)L}{2}, \frac{2 + 2\varepsilon - (1 - 2\varepsilon)L}{2}; \frac{3}{2}; \frac{(L+1)^{2}M^{2}}{t} \right).
\end{align}
When $L = 0$ we find
\begin{align}
	\mathcal{E}_{0}(t) &= \sqrt{2} \left( - \frac{M^{2}}{t} \right)^{(1-\varepsilon)} \frac{\Gamma\left( 1 - \varepsilon \right)}{\Gamma\left( \frac{1 - 2\varepsilon}{2} \right)} {}_{2} F_{1} \left( 1 - \varepsilon, \frac{1 + 2\varepsilon}{2}; \frac{1}{2}; \frac{M^{2}}{t} \right); \\
	\mathcal{O}_{0}(t) &= -\sqrt{8} \left( - \frac{M^{2}}{t} \right)^{(3 - 2\varepsilon)/2} \frac{\Gamma\left( \frac{3 - 2\varepsilon}{2} \right)}{\Gamma\left( - \varepsilon \right)} {}_{2} F_{1} \left( \frac{3 - 2\varepsilon}{2}, 1 + \varepsilon; \frac{3}{2}; \frac{M^{2}}{t} \right).
\end{align}
which are well-behaved in the $\varepsilon \rightarrow 0$ limit:
\begin{equation}
	\mathcal{E}_{0}(t) \rightarrow \sqrt{\frac{2}{\pi}}\frac{M^{2}}{M^{2} - t}, \qquad
	\mathcal{O}_{0}(t) \rightarrow 0.
\end{equation}
Indeed, we find the familiar tree-level contribution. For $L = 1$ and $\varepsilon \rightarrow 0$ we find
\begin{align}
	\mathcal{E}_{1}(t) &= \frac{\sqrt{\pi}}{4} \sqrt{- \frac{2M^{2}}{t}}; \\
	\mathcal{O}_{1}(t) &= \frac{1}{2\sqrt{2\pi}} \sqrt{\frac{4M^{2}}{t}} \operatorname{artanh}{\left( \sqrt{\frac{4M^{2}}{t}} \right)}.
\end{align}
However, starting with $L = 2$ we find a divergence near $\varepsilon = 0$:
\begin{align}
	\mathcal{E}_{2}(t) &= \frac{2^{(8\varepsilon - 3)/2}}{3} \left( - \frac{M^{2}}{t} \right)^{\varepsilon} \frac{\Gamma\left( \varepsilon \right)}{\Gamma\left( \frac{3 - 6\varepsilon}{2} \right)} {}_{2} F_{1} \left( \varepsilon, \frac{6 \varepsilon - 1}{2}; \frac{1}{2}; \frac{9M^{2}}{t} \right); \\
	\mathcal{O}_{2}(t) &= -2^{(8\varepsilon - 1)/2} \left( - \frac{M^{2}}{t} \right)^{(1 + 2\varepsilon)/2} \frac{\Gamma\left( \frac{1 + 2\varepsilon}{2} \right)}{\Gamma\left( 1 - 3 \varepsilon \right)} {}_{2} F_{1} \left( 3\varepsilon, \frac{1 + 2\varepsilon}{2}; \frac{3}{2}; \frac{9M^{2}}{t} \right).
\end{align}
Taking the $\varepsilon \rightarrow 0$ limit leads to
\begin{equation}
	\mathcal{O}_{2}(t) = - \frac{\sqrt{\pi}}{2} \sqrt{-\frac{2M^{2}}{t}};
\end{equation}
but $\mathcal{E}_{2}$ has a divergent part. Using the identity
\begin{equation}
	\Gamma(a) {}_{2}F_{1}(a, b; c; z) = \Gamma(a) + \frac{\Gamma(c)}{\Gamma(b)} \sum_{n = 1}^{\infty} \frac{\Gamma(a + n) \Gamma(b + n)}{\Gamma(c + n)} \frac{z^{n}}{n!};
\end{equation}
leads to
\begin{equation}
	\mathcal{E}_{2}(t) \approx \frac{1}{3 \sqrt{2 \pi}} \left[ \left(- \frac{2M^{2}}{t} \right)^{\varepsilon}\Gamma(\varepsilon) - 2 \sqrt{\frac{9M^{2}}{t}} \operatorname{artanh}{\left( \sqrt{\frac{9M^{2}}{t}} \right)} - \log{\left(1 - \frac{9 M^{2}}{t} \right)} \right].
\end{equation}
Similarly with $L = 3$: Taking the $\varepsilon \rightarrow 0$ limit leads to
\begin{equation}
	\mathcal{E}_{3}(t) = \sqrt{\pi} \sqrt{- \frac{2M^{2}}{t}} \left( 1 + \frac{t}{16M^{2}} \right);
\end{equation}
but $\mathcal{O}_{3}(t)$ has a divergent part,
\begin{equation}
\begin{split}
	\mathcal{O}_{3}(t) \approx \frac{1}{3\sqrt{2 \pi}} &{} \left[ \left(- \frac{2M^{2}}{t} \right)^{2\varepsilon}\Gamma(2\varepsilon) - \left( \sqrt{\frac{16 M^{2}}{t}} + \sqrt{\frac{t}{16 M^{2}}} \right) \operatorname{artanh}{\left( \sqrt{\frac{16M^{2}}{t}} \right)} \right. \\
	& \qquad - \left. \log{\left(1 - \frac{16 M^{2}}{t} \right)} + 1 \right].
\end{split}
\end{equation}
In general, for $L$ even and $L \geq 2$ we find that $\mathcal{E}_{L}(t)$ has a divergent part, and for $L$ odd and $L \geq 3$ we find that $\mathcal{O}_{L}(t)$ has a divergent part. Each of these divergent terms consist of a simple pole at $\varepsilon = 0$ (i.e. a term proportional to $\Gamma(n \varepsilon)$).
\section{Discussion}
Using the forward-JWKB approximation we obtained three sets of results: amplitudes in $D = 3$ with massless mediation, amplitudes in $D = 4$ with massless mediation, and amplitudes in $D = 3$ and $D = 5$ with massive mediation.

In $D = 4$ with massless mediation, we found forward-JWKB amplitudes (\ref{AHN4}) for mediating quanta with spin $0$, $1$ and $2$. Each of these amplitudes exhibits an infinite number of singularities that lie outside of the physical scattering region. Indeed, the singularities (\ref{sJ0}), (\ref{sJ1}) and (\ref{sJ2}) agree with the two-body bound-state energies found in \cite{BIZJ,KabatOrtiz,Dittrich}. Upon taking the static limit ($m_{1} / m_{2} \rightarrow 0$), the resulting one-body amplitudes agree with those found in \cite{Singh} by solving one-body field equations. All of these amplitudes display an exponentiated divergent contribution, the kinematic dependence of which agrees with those found in \cite{Weinberg:1965nx} due to infrared photons and gravitons. Inside the physical scattering region, the coefficient of this divergence is imaginary and thus the whole divergence appears as a pure phase factor. This guarantees that physical observables are finite.

\begin{figure}
\centering
\includegraphics[scale=0.56]{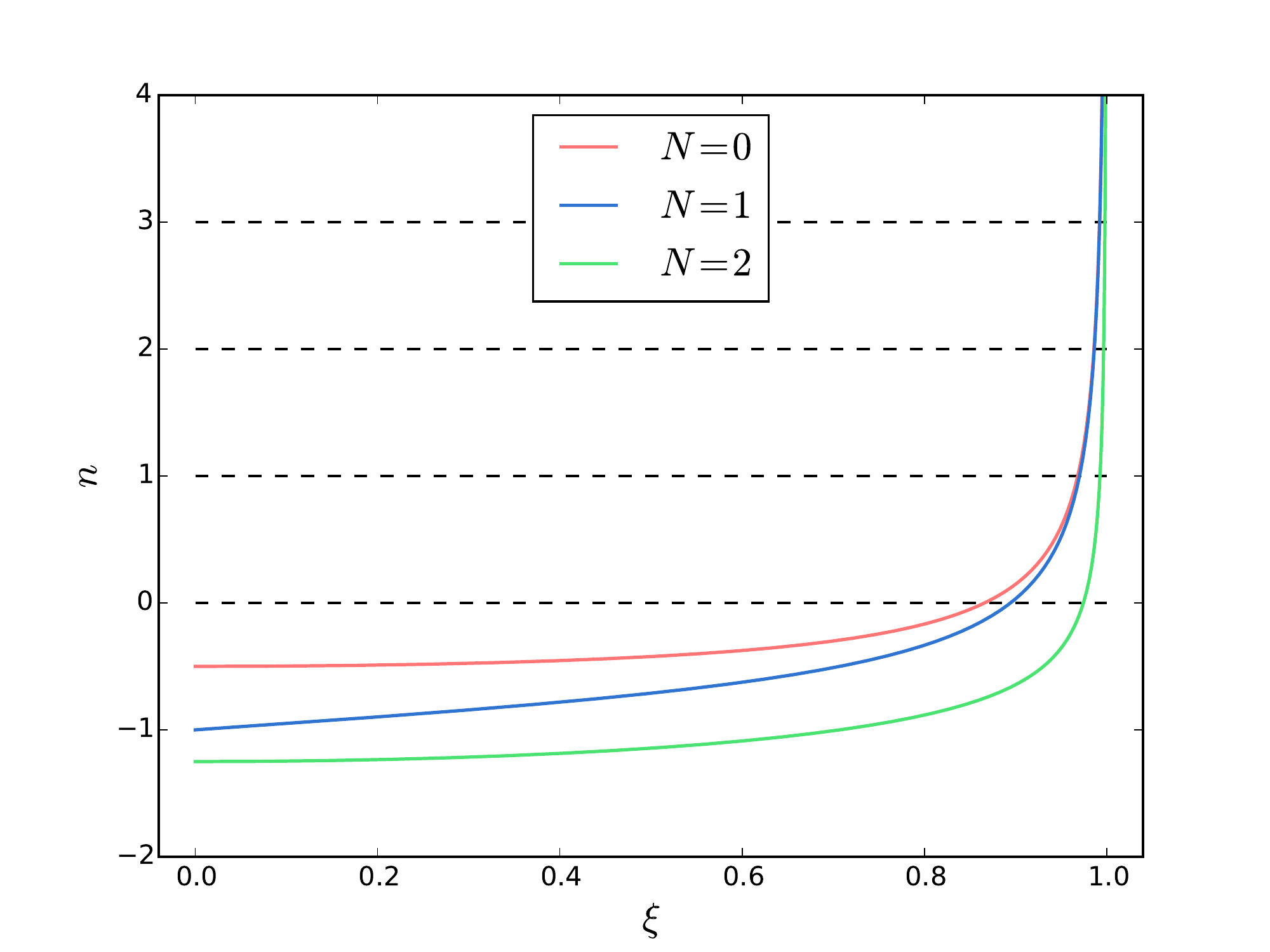}
\caption{The leading Regge trajectory functions $R_{N}(\xi)$ for spin $N = 0$, $1$ and $2$ in the forward-JWKB regime. The dashed lines correspond to integer values of $n$. For the couplings we have used $\alpha_{0} / (M_{1} M_{2}) = 0.5$, $Z_{1} Z_{2} \alpha_{1} = -0.5$, and $M_{1} M_{2} \alpha_{2} = 0.5$.}
\label{fig012}
\end{figure}

From the discussion in \S\ref{sec2} it should be clear that the Regge regime and the forward-JWKB regime are very different. The amplitude (\ref{AHN4}) in $D = 4$ exhibits Regge behavior with leading Regge trajectory function
\begin{equation}
	R_{N}(s) = -1 + \alpha_{N} \rho_{N}(s);
\end{equation}
and daughter trajectories $R_{N}(s) - l$. In Figure \ref{fig012} we plot the forward-JWKB leading Regge trajectories for $N = 0$, $1$ and $2$ as functions of the dimensionless variable
\begin{equation}
	\xi(s) \equiv \frac{s - M_{1}^{2} - M_{2}^{2}}{2 M_{1} M_{2}} \approx \frac{s - m_{1}^{2} - m_{2}^{2}}{2 m_{1} m_{2}}.
\end{equation}

\begin{figure}
\centering
\includegraphics[scale=0.56]{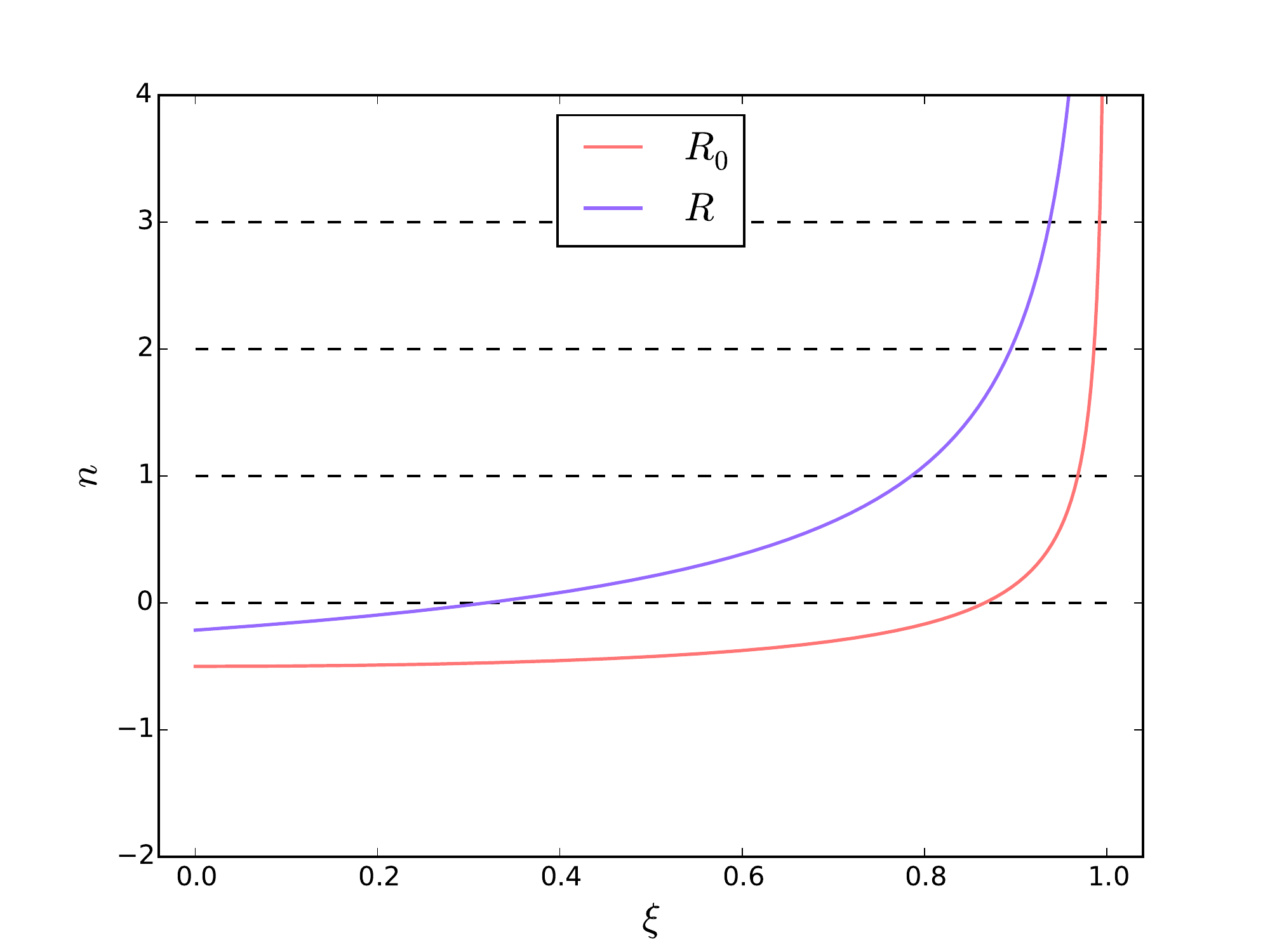}
\caption{The leading Regge trajectory for an interaction mediated by a massless scalar in the Regge limit ($R$) and the forward-JWKB regime ($R_{0}$).}
\label{figR0R}
\end{figure}

When $N = 0$ the forward-JWKB leading Regge trajectory $R_{0}$ is qualitatively different from the leading Regge trajectory function $R$ found in the Regge limit (\ref{ladders}). In Figure \ref{figR0R} we compare the leading trajectory in these two regimes. This comparison is only qualitative, as the normalization of the coupling strength $g$ in the Regge ladder sum (\ref{ladders}) is not necessarily the same as the one we used in the forward-JWKB amplitude. One major difference is that the real part of the Regge limit trajectory $R$ is non-vanishing in the region $\xi < 1$, while the real part of the forward-JWKB trajectory $R_{0}$ is non-vanishing in the smaller interval $-1 < \xi < 1$. However, near the threshold value both trajectories agree. Indeed, the discrepancy between these two trajectories was already noticed in \cite{Levy:1970yn}, where diagrammatic methods were used to sum over ladder contributions. Looking at the equations for $\rho$ in (\ref{LeeSawyer}) and $\rho_{0}$ in (\ref{rho0D4}) we see that the logarithm term in $\rho$ is missing from $\rho_{0}$. This logarithm term is proportional to the sum of the rapidities of the incoming states (\ref{rap12}), and is kept fixed in both the Regge and forward-JWKB limits. In future work we hope to explore this discrepancy further.

The forward-JWKB result (\ref{AHN4}) includes a factor involving an Euler Gamma function that is missing from the Regge limit ladder result (\ref{ladders}). A similar factor can be obtained using the Bethe-Salpeter equation \cite{LeeSawyer}. In order to obtain such a factor from field theory, one can use the Mellin transform technique \cite{BjorkenWu,TruemanYao} on the sum over ladder diagrams \cite{Polkinghorne1,Polkinghorne2,SMatrixBook} which helps to pick out non-leading logarithmic contributions.

The forward-JWKB amplitude in $D = 3$ with massless mediation exhibits the familiar tree-level singularity near $t = 0$, and also an ``extra'' non-perturbative singularity at a particular value $s = s_{*}(t_{*})$ that depends explicitly on $t_{*}$. For mediating scalars and photons, the location of the extra singularity is very similar to the location of the corresponding two-body bound state energies $s_{nl}$ in $D = 4$. Indeed, one can ``switch'' between the many singularities $s_{nl}$ and the one singularity $s_{*}$ with the replacement
\begin{equation}
	\frac{2 \pi \beta^{2}_{N}}{t_{*}} \longleftrightarrow \frac{\alpha^{2}_{N}}{(n + l + 1)^{2}};
	\label{74}
\end{equation}
where $\beta_{N}$ is the coupling in $D = 3$, and $\alpha_{N}$ is the coupling in $D = 4$. Upon analytic continuation to spin $2$, we also find a forward-JWKB amplitude with two singularities. However, unlike the scalar or photon cases, the $s_{*}$ singularity does not seem to correspond to anything that can be built with perturbative contributions (i.e. from contributions that are proportional to positive integer powers of the coupling). Unlike the $D = 4$ singularities $s_{nl}$, we seem to always be able to move the $D = 3$ singularity $s_{*}$ to the interior of the physical scattering region by allowing $t_{*}$ to be negative. Of course, the relation (\ref{74}) only holds when $t_{*}$ is positive.

We also studied the mediation of a heavy scalar in $D = 3$. In this case we found a non-perturbative expression that agrees with the expected result from adding forward-JWKB ladder diagrams of the form (\ref{fJWKBBox}) and (\ref{fJWKBDoubleBox}). Due to the particular details of the long-distance propagator in $D - 2 = 1$ dimensions, the result can be written as a sum of simple poles at the multi-mass values $(L+1)^{2}M^{2}$. The end result (\ref{615}) takes a very simple form that is analogous to Regge behavior, but instead of Regge poles along the $s$ axis, we find poles along the $t$ axis.

Then, instead of turning to $D = 4$ with a heavy mediator, we turned to $D = 5$. The reason for this is that when $D = 5$ we have $D - 2 = 3$ which is the other instance when the long-distance propagator is exact. Although we cannot evaluate the ``eikonal'' integral (\ref{DAvarphi}) exactly, after a series expansion we recover the familiar tree-level amplitude, a finite one-loop contribution and divergent higher-loop contributions. This system in $D = 5$ was considered mostly to illustrate that for higher-dimensional systems one can apply the forward-JWKB approximation and obtain limited results for all-order amplitudes.

In our calculation of forward-JWKB amplitudes we neglected the part of the forward Van Vleck matrix with derivatives of $\Sigma_{\text{two}}$ (the term with two-body interactions). The Van Vleck determinant is the first-order JWKB contribution. When written in the form (\ref{sqrtV}), it is clear that this contribution involves an infinite set of perturbative contributions (i.e. involving powers of the coupling). This demonstrates nicely the non-perturbative nature of the JWKB approximation. In principle, incorporating the contributions from $W$ in (\ref{sqrtV}) offers a way to go beyond the familiar $D = 4$ results obtained in this work. However, in practice it is not clear at the moment how to then evaluate the integrals in $\mathcal{S}_{T}$ and find the amplitude.

At first glance, the statement of the forward-JWKB approximation (\ref{fJWKBLimit}) amounts to a restriction on kinematics. But as was found in \cite{HalpernSiegel}, the semiclassical approximation has dynamical consequences that depend on the number of spacetime dimensions. Indeed, we found that in $D = 3$ the forward-JWKB approximation was consistent with weak couplings $\beta_{N}$, but in $D = 4$ we require strong couplings $\alpha_{N}$ when the quantum number becomes large. This strong-coupling feature is attractive and one hopes to be able to extend it to other theories (or at least to higher-point forward-JWKB scattering). It is also one of the reasons why we believe semiclassical first-quantized methods are important and useful to obtain non-perturbative results.
\ack{M.E.I.G. would like to acknowledge the generous support of the Turner Fellowship and the Center for Inclusive Education at Stony Brook University from 2006 to 2013, and the kind hospitality of the Physics Department at Wesleyan University during the 2013-2014 academic year. This work was supported in part by National Science Foundation Grant No. PHY-1316617.}
\appendix
\section{Four-Point Kinematics\label{app1}}
We study the scattering of two non-identical massive scalar particles. The external states are labeled such that the $s$-channel process is elastic:
\begin{equation}
	\Phi_{1}(p_{1}) + \Phi_{2}(p_{2}) \longrightarrow \Phi_{1}(p_{3}) + \Phi_{2}(p_{4}).
	\label{sChannel}
\end{equation}
Note that the $u$-channel process is also elastic:
\begin{equation}
	\Phi_{1}(p_{1}) + \bar{\Phi}_{2}(\bar{p}_{2}) \longrightarrow \Phi_{1}(p_{3}) + \bar{\Phi}_{2}(\bar{p}_{4});
	\label{uChannel}
\end{equation}
but the $t$-channel process is inelastic:
\begin{equation}
	\Phi_{1}(p_{1}) + \bar{\Phi}_{1}(\bar{p}_{2}) \longrightarrow \bar{\Phi}_{2}(\bar{p}_{3}) + \Phi_{2}(p_{4}).
	\label{tChannel}
\end{equation}
We will only consider the amplitude for the (\ref{sChannel}) process. The amplitude for the (\ref{uChannel}) process follows after setting $\bar{p}_{2} = -p_{4}$ and $\bar{p}_{4} = -p_{2}$, and the amplitude for the (\ref{tChannel}) process follows after setting $\bar{p}_{2} = -p_{3}$ and $\bar{p}_{3} = -p_{2}$.

The four external energy-momentum vectors satisfy the on-shell constraints,
\begin{equation}
	p_{j}^{2} = - m_{j}^{2};
\end{equation}
the elasticity constraints,
\begin{equation}
	p_{1}^{2} = p_{3}^{2}, \qquad p_{2}^{2} = p_{4}^{2} \quad \Longrightarrow \quad m_{1} = m_{3}, \qquad m_{2} = m_{4}; \label{elas}
\end{equation}
and also the conservation constraint,
\begin{equation}
	p_{1} + p_{2} = p_{1} + p_{4}.
\end{equation}
We use the familiar Mandelstam energy-momentum invariants
\begin{equation}
	s = -(p_{1} + p_{2})^{2}, \qquad t = -(p_{1} - p_{3})^{2}, \qquad u = -(p_{1} - p_{4})^{2};
\end{equation}
which satisfy
\begin{equation}
	s + t + u = 2 m_{1}^{2} + 2 m_{2}^{2}.
\end{equation}
Note that $s$ and $u$ carry data from both particles, while $t$ carries data from a single particle.

In the center-of-momentum frame, we write the energy-momentum vectors as
\begin{equation}
	p_{1} = \begin{pmatrix} E_{1} & \mathbf{p}_{1} \end{pmatrix}, \qquad p_{2} = \begin{pmatrix} E_{2} & -\mathbf{p}_{1} \end{pmatrix}, \qquad p_{3} = \begin{pmatrix} E_{3} & \mathbf{p}_{3} \end{pmatrix}, \qquad p_{4} = \begin{pmatrix} E_{4} & -\mathbf{p}_{3} \end{pmatrix}.
\end{equation}
It is easy to show that, in terms of $s$ and the masses $(m_{1}, m_{2})$, the magnitude of the spatial vectors $(\mathbf{p}_{1}, \mathbf{p}_{3})$ are given by
\begin{equation}
	|\mathbf{p}_{1}| = |\mathbf{p}_{3}| = \frac{\sqrt{\Lambda_{12}}}{2 \sqrt{s}}, \qquad \Lambda_{12} \equiv \left[s - \left(m_{1} - m_{2} \right)^{2} \right] \left[s - \left(m_{1} + m_{2}\right)^{2} \right];
	\label{2Momenta}
\end{equation}
and the energies of the external states are given by
\begin{equation}
	E_{1} = E_{3} = \frac{s + m_{1}^{2} - m_{2}^{2}}{2 \sqrt{s}}, \qquad E_{2} = E_{4} = \frac{s - m_{1}^{2} + m_{2}^{2}}{2 \sqrt{s}}.
	\label{4Energies}
\end{equation}
A relativistic particle with mass $m \geq 0$ and energy $E \geq m$ has a speed $|\mathbf{v}| \leq 1$ given by
\begin{equation}
	|\mathbf{v}| = \frac{\sqrt{E^{2} - m^{2}}}{E}.
\end{equation}
Thus, in the center-of-momentum frame the external states have speeds,
\begin{equation}
\begin{split}
	|\mathbf{v}_{1}| &= |\mathbf{v}_{3}| = \frac{\sqrt{\Lambda_{12}}}{s + m_{1}^{2} - m_{2}^{2}}, \\
	|\mathbf{v}_{2}| &= |\mathbf{v}_{4}| = \frac{\sqrt{\Lambda_{12}}}{s - m_{1}^{2} + m_{2}^{2}};
\end{split} \label{4Speeds}
\end{equation}
and rapidities $\varphi_{j} \equiv \operatorname{artanh}{(|\mathbf{v}_{j}|)}$,
\begin{equation}
\begin{split}
	\varphi_{1} &= \varphi_{3} = \frac{1}{2} \log{\left[ \frac{s + m_{1}^{2} - m_{2}^{2} + \sqrt{\Lambda_{12}}}{s + m_{1}^{2} - m_{2}^{2} - \sqrt{\Lambda_{12}}} \right]}, \\
	\varphi_{2} &= \varphi_{4} = \frac{1}{2} \log{\left[ \frac{s - m_{1}^{2} + m_{2}^{2} + \sqrt{\Lambda_{12}}}{s - m_{1}^{2} + m_{2}^{2} - \sqrt{\Lambda_{12}}} \right]}.
\end{split} \label{4Rapidities}
\end{equation}
Note that the sum of the incoming rapidities gives
\begin{equation}
	\varphi_{1} + \varphi_{2} = \frac{1}{2} \log{\left[ \frac{s - m_{1}^{2} - m_{2}^{2} + \sqrt{\Lambda_{12}}}{s - m_{1}^{2} - m_{2}^{2} - \sqrt{\Lambda_{12}}} \right]}.
\end{equation}
Using
\begin{equation}
	\frac{s - (m_{1} + m_{2})^{2} + \sqrt{\Lambda_{12}}}{s - (m_{1} + m_{2})^{2} - \sqrt{\Lambda_{12}}} = -\frac{2m_{1}m_{2}}{s - m_{1}^{2} - m_{2}^{2} - \sqrt{\Lambda_{12}}} = -\frac{s - m_{1}^{2} - m_{2}^{2} + \sqrt{\Lambda_{12}}}{2m_{1}m_{2}};
\end{equation}
we can also write
\begin{equation}
	\varphi_{1} + \varphi_{2} = \log{\left[ \frac{s - (m_{1} + m_{2})^{2} + \sqrt{\Lambda_{12}}}{s - (m_{1} + m_{2})^{2} - \sqrt{\Lambda_{12}}} \right]}.
	\label{rap12}
\end{equation}
The cosine of the angle between $\mathbf{p}_{1}$ and $\mathbf{p}_{3}$ is given by
\begin{equation}
	z_{s} \equiv \cos(\theta_{s}) = \frac{\mathbf{p}_{1} \cdot \mathbf{p}_{3}}{|\mathbf{p}_{1}| |\mathbf{p}_{3}|} = \frac{(m_{1} - m_{2})^{2} (m_{1} + m_{2})^{2} - s (u - t)}{(m_{1} - m_{2})^{2} (m_{1} + m_{2})^{2} - s (u + t)}.
	\label{z13rational}
\end{equation}
The angle $\theta_{s}$ is known as the scattering angle.
\subsection{Physical Scattering Region}
Requiring each of the energies in (\ref{4Energies}) to be real and non-negative leads to the condition
\begin{equation}
	s \geq |m_{1} - m_{2}| (m_{1} + m_{2}).
\end{equation}
Indeed, the same requirements on the magnitudes in (\ref{2Momenta}) leads to a stronger condition:
\begin{equation}
	\Lambda_{12} \geq 0 \quad \Longrightarrow \quad s \geq (m_{1} + m_{2})^{2} > |m_{1} - m_{2}| (m_{1} + m_{2}).
\end{equation}
We must also require $z_{s}$ in (\ref{z13rational}) to satisfy
\begin{equation}
	{-1} \leq z_{s} \leq 1.
\end{equation}
This is equivalent to the conditions
\begin{equation}
	t \leq 0, \qquad s u \leq (m_{1} - m_{2})^{2} (m_{1} + m_{2})^{2}.
\end{equation}
Thus, the physical scattering region is defined by
\begin{equation}
	s \geq (m_{1} + m_{2})^{2}, \qquad t \leq 0, \qquad s u \leq (m_{1} - m_{2})^{2} (m_{1} + m_{2})^{2}.
\end{equation}
Since these relations involve Lorentz invariants, they hold on any reference frame related to the center-of-momentum frame by a Lorentz transformation.
\bibliography{forward4Bib}
\end{document}